\documentclass[preprint]{aastex}

\usepackage{natbib}
\usepackage{amsfonts}
\usepackage{pdflscape}
\usepackage{rotating}
\usepackage[normalem]{ulem}

\shorttitle{Testing Magnetic Field Models for the Class 0 Protostar L1527}
\shortauthors{Davidson et al.}

\begin{document}

\title{Testing Magnetic Field Models for the Class 0 Protostar L1527}

\author{J.A. Davidson\altaffilmark{1}, Z-Y. Li\altaffilmark{2}, C.L.H. Hull\altaffilmark{3,13,14}, R.L. Plambeck altaffilmark{3}, W. Kwon\altaffilmark{4}, R.M. Crutcher, and L.W. Looney\altaffilmark{5}, G. Novak and N.L. Chapman\altaffilmark{6}, B.C. Matthews\altaffilmark{7}$^,$\altaffilmark{8}, I.W. Stephens\altaffilmark{9}, J.J. Tobin\altaffilmark{10}$^,$\altaffilmark{12}, and T.J. Jones\altaffilmark{11}\\
\textit{(Submitted to ApJ on 24 August 2014; accepted 4 October 2014.)}}

\altaffiltext{1}{University of Western Australia, School of Physics, 35 Stirling Hwy, Crawley, WA 6009, Australia; jackie.davidson@uwa.edu.au}
\altaffiltext{2}{University of Virginia, Astronomy Department, Charlottesville, VA 22904, USA}
\altaffiltext{3}{University of California, Astronomy Department \& Radio Astronomy Laboratory, Berkeley, CA 94720-3411, USA}
\altaffiltext{4}{SRON Netherlands Institute for Space Research, Landleven 12, 9747 AD, Groningen, The Netherlands}
\altaffiltext{5}{University of Illinois, Department of Astronomy, 1002 West Green St, Urbana, IL 61801, USA}
\altaffiltext{6}{Northwestern University, Center for Interdisciplinary Exploration and Research in Astrophysics (CIERA) and the Department of Physics \& Astronomy, 2145 Sheridan Road, Evanston, IL 60208, USA} 
\altaffiltext{7}{Herzberg Astronomy \& Astrophysics, National Research Council of Canada, 5071 West Saanich Road, Victoria, BC, V9E 2E7, Canada}
\altaffiltext{8}{University of Victoria, Department of Physics \& Astronomy, 3800 Finnerty Road, Victoria, BC, V8P 1A1, Canada} 
\altaffiltext{9}{Boston University, Institute for Astrophysical Research, Boston, MA 02215, USA}
\altaffiltext{10}{National Radio Astronomy Observatory, 520 Edgemont Rd., Charlottesville, VA 22903, USA}
\altaffiltext{11}{University of Minnesota, 116 Church St S.E., Minneapolis, MN, 55455 USA}
\altaffiltext{12}{Hubble Fellow}
\altaffiltext{13}{Harvard-Smithsonian Center for Astrophysics, 60 Garden St, MS 42, Cambridge, MA 02138, USA}
\altaffiltext{14}{Jansky Fellow of the National Radio Astronomy Observatory, which is a facility of the National Science Foundation operated under cooperative agreement by Associated Universities, Inc.}

\begin{abstract}
For the Class 0 protostar, L1527, we compare 131 polarization vectors from SCUPOL/JCMT, SHARP/CSO and TADPOL/CARMA observations with the corresponding model polarization vectors of four ideal-MHD, non-turbulent, cloud core collapse models.  These four models differ by their initial magnetic fields before collapse; two initially have aligned fields (strong and weak) and two initially have orthogonal fields (strong and weak) with respect to the rotation axis of the L1527 core.  Only the initial weak orthogonal field model produces the observed circumstellar disk within L1527. This is a characteristic of nearly all ideal-MHD, non-turbulent, core collapse models. In this paper we test whether this weak orthogonal model also has the best agreement between its magnetic field structure and that inferred from the polarimetry observations of L1527.  We found that this is not the case; based on the polarimetry observations the most favored model of the four is the weak aligned model.  However, this model does not produce a circumstellar disk, so our result implies that a non-turbulent, ideal-MHD global collapse model probably does not represent the core collapse that has occurred in L1527. Our study also illustrates the importance of using polarization vectors covering a large area of a cloud core to determine the initial magnetic field orientation before collapse; the inner core magnetic field structure can be highly altered by a collapse and so measurements from this region alone can give unreliable estimates of the initial field configuration before collapse. 
 
\end{abstract}


\section{Introduction}

Magnetic fields are thought to play a crucial role in the star formation process (e.g., McKee \& Ostriker 2007). The radius and rotation periods of classical T Tauri stars measured by Bouvier et al. (1993) and the cloud core rotations measured by Goodman et al. (1993) on the scale of 0.1 pc (20,000 AU), imply a specific angular momentum reduction of about 5 orders of magnitude by the cloud core material as it forms a star (Tomisaka, 2000). Furthermore, an additional reduction in specific angular momentum of about two orders of magnitude is required by the material forming the initial cloud core from the diffuse ISM, if angular momentum is conserved by the material initially co-rotating with the galactic background (Basu \& Mouschovias, 1994). Magnetic braking is a commonly proposed mechanism to achieve reduction of angular momentum (e.g., Spitzer, 1978).  Mouschovias \& Paleologou (1979) showed that a uniform field is more efficient at reducing specific angular momentum if it is perpendicular to the axis of rotation rather than aligned with it.  This would suggest that the component of cloud rotation perpendicular to the magnetic field will be more efficiently suppressed than the component of rotation that is aligned with the magnetic field.  The result would be a cloud core rotation axis that is more aligned than perpendicular to the magnetic field in that cloud core prior to collapse.  Many studies have indeed assumed an aligned magnetic field with the rotation axis when numerically or analytically studying the subsequent collapse within these cloud cores to form protostars (e.g., Tomisaka 1998;  Allen et al. 2003; Galli et al. 2006).  

However, recent ideal-MHD theoretical studies of non-turbulent cloud core collapse in star forming regions indicate that the formation of rotationally supported circumstellar disks is suppressed if the initial cloud core magnetic field is relatively strong and the magnetic field is more aligned to the rotation axis of the cloud core than orthogonal to it (Hennebelle \& Ciardi, 2009; Joos et al., 2012;  Li et al., 2013). This conclusion may not hold if non-ideal MHD effects, such as Ohmic dissipation (Dapp \& Basu 2010), and/or turbulence (Santos-Lima et al. 2012) are incorporated.  Observational evidence strongly favours the existence of circumstellar disks around more evolved young stellar objects (YSOs) (see e.g., J\o rgensen et al. 2009 and Williams \& Cieza 2011 and references therein), but there are relatively few examples of detected circumstellar disks surrounding Class 0 YSOs.  This is unfortunate, since it is this class of YSOs which still has a large fraction of their envelopes intact enabling studies of the initial conditions of the cloud core before collapse.  Polarimetry studies of this small subset of Class 0 YSOs (with detected circumstellar disks) would test the hypothesis of the ideal-MHD models in regards to the orientation of their initial magnetic fields with respect to the rotation axis of the YSO's cloud core.  Such studies would assume dust grain alignment as outlined in Lazarian (2007), where the polarization by dust emission is perpendicular to the direction of the projected magnetic field on the plane of the sky at the location of the emission. 

In this paper, we focus on one of the most well-studied low-mass, Class 0 protostars, L1527, the first Class 0 YSO with an observed Keplerian disk (Tobin et al. 2012).  Over one hundred polarization vectors have been measured in the cloud core of L1527 (Matthews et al., 2009; Davidson et al., 2011; Hull et al., 2014), covering size-scales of 20,000 AU down to 400 AU, enabling us to test cloud core collapse models in detail in this one source.  In this paper we will test four ideal-MHD core collapse models of the same type as those that predict the suppression of circumstellar disks if the field is more aligned than orthogonal to the rotation axis of the collapsing cloud core. We will determine if indeed the observed magnetic field direction inferred from the polarimetry observations of L1527 favours an initial orthogonal magnetic field to the cloud's rotation axis.   This detailed approach complements the survey approach (e.g., Krumholz et al., 2013; Hull et al., 2013; Hull et al., 2014), which uses spatial averages of sometimes limited samples of polarization vectors to determine the overall initial magnetic field configuration for each source.  We will examine the type of averaging method used in these surveys to evaluate their accuracy in determining the initial cloud core magnetic field configuration.

Section 2 below summarizes the observational data obtained for L1527 and the cloud core collapse models we are using to simulate the collapse in L1527.  This section also outlines the approach we have taken to simulate the measured L1527 polarization data using the models. Section 3 summarizes our results using various $\chi^2$ statistics to compare the modelled data with the real data for L1527.  In Section 4 we discuss the results of Section 3 and compare our results to surveys and other polarimetry studies for this source and similar sources.  Our conclusions are given in Section 5.

\section{Observations and Models}

\subsection{L1527 Observations}

L1527 lies within the Taurus cloud complex at a distance of about 140 pc (Kenyon et al. 1994; Torres et al. 2009).  The near-IR and mid-IR observations of Tobin et al. (2008) constrain the outer radius of the cloud core surrounding the L1527 YSO to be 15000 AU. The total mass of this cloud core is $\sim$ 2 M$_{\sun}$ as estimated by the modelling of the near-IR, mid-IR, far-IR, and sub-MM continuum images of L1527 (Tobin et al., 2008) and  the NH$_3$ line observations of this source by Benson \& Myers (1989).   Ohashi et al. (1997) observed in the C$^{18}$O (J=1$\rightarrow$0) line a north-south flattened column density enhancement with a radius of $\sim$ 2000 AU.  In addition, Ohashi et al. interpreted the line profiles of C$^{18}$O as infall and rotation within this flattened inner cloud core structure.  Their results indicate the structure is not rotationally supported.   The rotation sense and magnitude measured by Ohashi et al. (1997) on the scale of 2000 AU ($\sim$ 5 kms$^{-1}$ pc$^{-1}$ pointing east) are consistent with the rotation measured by Tobin et al. (2011) inside a radius of $\sim$2000 AU using the N$_2$H$^+$(1-0) and NH$_3$(1,1) lines.  At larger scales in the cloud core, Tobin et al. (2011) found a more complex velocity structure, which is possibly a combination of infall, rotation and turbulence.  The molecular line observations of Myers et al. (1995) and Zhou et al. (1996) indicate the possible existence of an infall radius at $\sim$ 5000 AU consistent with the Shu (1977) inside-out collapse model (more on this in Section 2.2).  L1527 also has a well defined east-west bipolar outflow (Zhou et al., 1996; Hogerheijde et al., 1998) which lies within 5$\arcdeg$ of the plane of the sky (Tobin et al., 2010). The outflow axis coincides with the rotation axis of the inner cloud core  of L1527 (i.e., within a radius of 2000 AU).  The geometry of the outflow and the measured line profile velocities imply the age for the outflow is only a few thousand years (Hogerheijde et al., 1998).  The near-IR and mid-IR scattered light analysis of Tobin et al. (2010) indicates the existence of two cavities carved out by the bipolar outflow.  These cavities, one to the east and one to the west of the YSO, are approximately conical in shape with their apexes at the position of the YSO and with half-opening angles of $\sim$ 20$\arcdeg$. Finally, the continuum and $^{13}$CO (J=2$\rightarrow$1) line interferometric observations of Tobin et al. (2012) reveal a rotationally supported (i.e., Keplerian) circumstellar disk with a radius of about 100 AU and a rotation axis pointing east, aligned with the outflow axis. This is one of the best examples of such a disk about a Class 0 YSO to date. The Keplerian motion of the material in this disk implies the mass of the central L1527 YSO to be about 0.2 M$_{\sun}$.

In addition to the above studies, there have been a number of L1527 polarimetry studies.  In this paper we are comparing our models with the polarization measurements made in three ``observation sets'':\\
\indent(1) The 1.3mm CARMA observation set containing 9 polarization vectors constructed with an effective 3$\arcsec$ FWHM beam on a Nyquist-sampled 1.5$\arcsec$ spatial grid (Hull et al., 2014).\\
\indent(2) The re-reduced 350$\mu$m  SHARP observation set containing 30 polarization vectors constructed with an effective 10$\arcsec$ FWHM beam on a Nyquist-sampled 5$\arcsec$ spatial grid.  These vectors are based on the same data reported in Davidson et al. (2011), but the data have been re-reduced as per Chapman et al. (2013) and re-binned in a Nyquist-sampled spatial grid rather than a beam-sampled grid.\\
\indent(3) The 850$\mu$m  SCUPOL observation set containing 92 polarization vectors constructed with an effective 20$\arcsec$ FWHM beam on a Nyquist-sampled 10$\arcsec$ spatial grid (Matthews et al., 2009).\\
The results of the re-reduced SHARP data are summarized in Table 1 of this paper.  The CARMA results are summarized in the on-line table of Hull et al. (2014), and the SCUPOL results are summarized in the on-line table of Matthews et al. (2009).  The orientations of the polarization-vectors reported in these tables are east from north.  These must be rotated by 90$^{\arcdeg}$ to show the inferred  magnetic field vectors (i.e., ``B-vectors''), since dust grains align with their long axes orthogonal to the magnetic field (Lazarian 2007).

Figures 1, 2 and 3 display the inferred B-vectors for the SCUPOL, SHARP, and CARMA observation sets used in this paper.  The length of each B-vector indicates the magnitude of the polarization fraction, $P$,  observed for the polarization-vector associated with that B-vector.  The black vectors have $P/\sigma_{_P} \ge  3$, while the grey vectors have $2 \le P/\sigma_{_P} <  3$, where $\sigma_{_P}$ are the 1-sigma errors of the polarization fractions.

\subsection{Cloud Core Collapse Models}

In this paper, we are comparing the polarimetry data of L1527 to four cloud core collapse models with differing initial magnetic field strengths and orientations.  We are using these models because they are similar to the ideal-MHD, ``global'' cloud core collapse models that predict the suppression of circumstellar disk formation when the initial magnetic fields are aligned with the rotation axis of the cloud core.  The initial physical configurations of these models are not initially in force balance, resulting in a global infall motion throughout the cloud core after the initial start time of the simulation.  This differs from an initial force balanced cloud core which has a singular, isothermal, spherical density structure as per Shu (1977).  In this latter case, the collapse region is bound by an expansion wave; there is collapse/infall inside this expansion wave, but the envelope remains static outside. This type of collapse is often referred to as an inside-out collapse.  As mentioned in Section 2.1, there is evidence of such an expansion wave in L1527, but this evidence does not conclusively eliminate a global collapse scenario for L1527 since the infall in a global collapse model is faster in the central regions than the outer regions of the core, and so the reported observations of an infall envelope may in fact be reflecting this difference in infall rates.

All four of the models employed in this paper satisfy the following three conditions, which are based on the observations:\\
\indent(1) The initial cloud core has a uniform density, a radius of 15,000 AU and contains a mass of 2 M$_{\sun}$.   This outer radius was set by scattered light observations (Tobin et al., 2008, 2010). The mass is consistent with Tobin et al. (2010) and Benson \& Myers (1989).\\
\indent(2) The initial solid-body rotation of this cloud core has a value of 2 kms$^{-1}$pc$^{-1}$ pointing east. This is consistent with an upper limit set by observations which show that the outer region of the L1527 cloud core has a complex velocity structure, some of which is not attributed to rotation (e.g., Tobin et al. 2011). The direction of the rotation is consistent with the better defined rotation observed inside a radius of 2000 AU (Tobin et al. 2011 and Ohashi et al. 1997), and the orientation of the observed E-W outflow in L1527 (Zhou et al., 1996; Hogerheijde et al. 1998).  \\
\indent(3) The epoch of our observations is set by the existence of a central 0.2 M$_{\sun}$  protostellar object (Tobin et al. 2012).  In other words, we ran the simulations for each of our four models described below until they produced a central 0.2  M$_{\sun}$ protostellar object.  The objective of this study is to compare the magnetic field structure at this epoch for each of these four models to that observed via the polarimetry observations of L1527.

The four global core collapse models we are studying can be identified by the strength and orientation of their initial uniform magnetic fields, where the strength is encapsulated in the dimensionless mass-to-(magnetic)flux ratio, $\lambda$ (Crutcher, 1999). $\lambda$ compares the mass-to-flux ratio of a given cloud core to its critical mass-to-flux ratio, corresponding to the value where a magnetized cloud core becomes unstable to gravitational collapse as defined by Nakano \& Nakamura (1978).  This critical mass-to-flux ratio assumes all other stabilising forces are insignificant, and so $\lambda$ can be greater than one in a stable cloud core if other forces such as rotation, turbulence and gas pressure are significant.  The OH Zeeman observations of 34 dark clouds by Troland \& Crutcher (2008) show that gravitational, magnetic and turbulent energies appear comparable and that $\lambda$ is approximately a few on average.  Initially, we ran our global collapse simulations for models having $\lambda =$ 3, but found that these simulations could not converge to form a central protostar of 0.2 M$_{\sun}$.  We reduced the magnetic field strength to $\lambda = 4$ and found that this enabled a central 0.2 M$_{\sun}$ to form for these models.  We call these models the ``strong'' field models.  Although these produced the required protostar mass, they could not reproduce the rotationally supported circumstellar disk observed by Tobin et al. (2012).  This is consistent with the results of Li et al. (2013), which show that a $\lambda$ value of about 10 is required for such models before Keplerian disks emerge (in the absence of any turbulence). We call these models the ``weak'' field models.
 
Thus the four models that were run for our L1527 comparison satisfy the above three conditions and can be characterized by their \underline{initial} uniform magnetic field strength and orientations as follows:\\
	\indent SA Model:  \underline{Strong} ($\lambda =$ 4) field, \underline{aligned} with the rotation axis of the cloud core;\\
	\indent WA Model:  \underline{Weak} ($\lambda = $ 10) field, \underline{aligned} with the rotation axis of the cloud core;\\
	\indent SO Model:  \underline{Strong} ($\lambda =$ 4)  field, \underline{orthogonal} to the rotation axis of the cloud core;\\
	\indent WO Model:  \underline{Weak} ($\lambda =$ 10) field, \underline{orthogonal} to the rotation axis of the cloud core. \\	
(See Appendix A for the technical details in regards to the model simulations.)

It is important to note that although all the models produce a central 0.2 M$_{\sun}$ protostar, only the WO model produces the rotationally supported disk observed by Tobin et al. (2012) for L1527; magnetic braking seems to have been too efficient in the other models.  In this paper we will test whether or not this WO model also has the best agreement, compared with our other models, between its modelled magnetic field structure and that inferred from the polarimetry observations of L1527.  

The outputs of the four model simulations are in spherical coordinates.  We have transformed these coordinates to sky RA and Dec ($\alpha$, $\delta$) and line of sight (z) coordinates to obtain:
 \[
 { \rho_{gas}(\alpha, \delta, z); \ \  B_{\alpha}(\alpha, \delta, z); \ \ B_{\delta}(\alpha, \delta, z); \ \ B_{z}(\alpha, \delta, z)}
 \]
 \begin{equation}
 {\xi_{_B}(\alpha, \delta, z) = \arctan \biggl({B_{\alpha} \over {B_{\delta}}}\biggr)}
 \end{equation}
 \begin{equation} 
 {\gamma(\alpha, \delta, z) = \arctan \biggl({B_{z} \over {\sqrt{B_{\alpha}^2 + B_{\delta}^2}}}\biggr)}
 \end{equation}
where $\rho_{gas}$ is the 3-D gas mass density, $\vec{B}= (B_{\alpha}, B_{\delta}, B_z)$ is the magnetic field vector, $\xi_{_B}$ is the east from north angle of the magnetic field projected on the plane of the sky, and $\gamma$ is the inclination angle of the magnetic field to the plane of the sky at $(\alpha, \delta, z)$.  For each simulation the $\alpha$-axis is uniquely defined by the rotation vector observed for L1527; this is the polar axis in the simulations.  However, the other two sky axes ($\delta$ and z) are not uniquely defined, and could be any two orthogonal axes lying in the plane perpendicular to the polar axis.  For each simulation, we have defined two possible lines of sight, LOS-1 and LOS-2, as shown in Figure 4, since we are not assuming axial symmetry about the rotation axis. This symmetry certainly cannot be assumed for the SO and WO models.  LOS-1 has been chosen to be the axis coinciding with the direction of the initial uniform magnetic field lines in the SO and WO models. Once a line of sight has been defined, the remaining sky axis (the $\delta$ axis) is uniquely defined.  

Figure 4 shows a  3-D schematic of the magnetic field lines for an initial aligned field model (a) and for an initial orthogonal field model (b), shortly after the initiation of the cloud core collapse.   The field lines in the aligned model for the most part maintain a pinched, hourglass, poloidal appearance, but at later stages of the collapse, and especially for the weaker fields, considerable toroidal twisting of the field lines occur in the core's center and to some degree along the axis of rotation.  The field lines in the orthogonal model exhibit a pinch with a twist, resulting in a strong toroidal configuration at the center of the cloud core, but also a surprisingly similar poloidal configuration in projection on the sky to the aligned model on spatial scales farther out from the cloud core's center.  This will be discussed further in Section 3.

As mentioned above, the observations of Tobin et al. (2010) indicate the existence of two cavities carved out by the bipolar outflow in L1527.  These cavities are approximately conical in shape with their apexes at the position of the YSO and with half-opening angles of $\sim$ 20$^{\arcdeg}$.   We have assumed, for modelling purposes, that the symmetry axis of the outflow cavities lies in the plane of the sky (true on the order of about 5$^{\arcdeg}$ for L1527) along the E-W axis (see Figure 5).  We have modelled the cavities by setting $\rho_{_{gas}}$ to zero within these cavities; so along each line of sight on the sky,  $\rho_{_{gas}}(\alpha, \delta, z) = 0$ for $-z_o < z < z_o$ where:
 \begin{equation}
 { z_o (\alpha, \delta) =  \sqrt{\alpha^2D^2 \tan^2 \theta_o - \delta^2 D^2} \ . }
 \end{equation}
and where $\theta_o$ =20$^{\arcdeg}$ and $D$ is the distance to L1527.

\subsection {B-Vector Models}

The Stokes parameters of the dust emission for each infinitesimal line of sight length element, $dz$, at ($\alpha, \delta, z$), can be given by:
\begin{equation}
{dI_{\nu}(\alpha, \delta, z) = B_{\nu}(T) d\tau_{\nu}}
\end{equation}
\begin{equation}
{dQ_{\nu}(\alpha, \delta, z) =  P_{\nu}B_{\nu}(T) \cos(2\xi) d\tau_{\nu}}
\end{equation}
\begin{equation}
{dU_{\nu}(\alpha, \delta, z) = P_{\nu}B_{\nu}(T) \sin (2\xi) d\tau_{\nu}}
\end{equation}
where $P_{\nu}$ is the degree of polarization of the emission (i.e., polarization fraction) from $dz$ at $(\alpha, \delta, z)$; $B_{\nu}(T)$ is the Planck function evaluated for a dust temperature of $T$ at the spatial position of $(\alpha, \delta, z)$; $d\tau_{\nu}$ is the optical depth along the infinitesimal portion of the line of sight $dz$ at $(\alpha, \delta, z)$; and $\xi$ is the linear polarization angle measured east from north for the sub-mm and mm emission at $(\alpha, \delta, z)$ (see Figure 9 in Hildebrand et al., 2000).  Since it is expected that dust polarization is orthogonal to the magnetic field orientation (due to dust grains aligning with their long axes orthogonal to the local magnetic field), we will assume $|\xi - \xi_{_B}| = 90\arcdeg$, where $\xi_{_B}$ is the east from north angle of the magnetic field projected onto the plane of the sky given by our models.

The polarization fraction at each location in 3-D space,  $P_{\nu}$, can be separated into two factors, $P^{ga}_{\nu}$ and $P^{mag}$.  The first is dependent on the shape and composition of the dust grains and the degree of the grain spin-axis alignment with the magnetic field at that location; the second is dependent on the inclination angle of the magnetic field to the plane of the sky, $\gamma$, at that location (Hildebrand \& Dragovan 1995).  Lee \& Draine (1985) showed that this second term equals $\cos^2 \gamma$ (see also Hildebrand 1988).  For this paper, we assume $P^{ga}_{\nu}$ is spatially constant throughout the cloud core, but we allow it to differ in value from one observing frequency to another. 

In addition, we further expand $d\tau_{\nu}$ in terms of dz:
\begin{equation}
{d\tau_{\nu}  = \rho_{_{dust}}  \kappa_{\nu} dz =  {\rho_{_{gas}} \over {100}}  \kappa_{\nu} dz, }
\end{equation}
where $\kappa_{\nu}$ is the dust opacity per unit dust mass, and $\rho_{_{dust}}$ and $\rho_{_{gas}}$ are the mass densities of the dust and gas, respectively, for a gas-to-dust mass ratio of 100.

In an optically thin cloud, the Stokes parameters given in (4), (5) and (6) can be summed along the line of sight for each position on the sky to give the total I, Q and U sky maps (Hildebrand et al., 2000).  However, the central regions of the L1527 models outlined above are not optically thin even for the long wavelengths being considered in this paper.  Appendix B shows how we model the transfer of linearly polarized radiation at the three wavelengths of our observation sets along each line of sight on the sky for each of our four cloud core collapse models to obtain the model Stokes parameters $I_{\nu}^{beam}(\alpha, \delta)$, $Q_{\nu}^{beam}(\alpha, \delta)$, and $U_{\nu}^{beam}(\alpha, \delta)$ as seen by an observer after beam effects have been included for a given observation set. 

The model Stokes parameters derived in Appendix B depend on wavelength (i.e., frequency $\nu$) because of the wavelength dependent quantities ($\kappa_{\nu}$, P$^{ga}_{\nu}$ and B$_{\nu}$(T)) included in their calculation.   For $\kappa_{\nu}$, we have used  1.11 cm$^2$/g, 10 cm$^2$/g, and 1.8 cm$^2$/g for the CARMA, SHARP and SCUPOL observation sets, respectively, based on columns 5 (SHARP and SCUPOL) and 6 (CARMA) of Table 1 of Ossenkopf \& Henning (1994). We have used column 6 values for the CARMA model since the CARMA observation set covers the densest region of the L1527 core where $\rho \sim 10^{-16}$ g/cm$^3$ and greater.  

P$^{ga}_{\nu}$, although assumed spatially constant in this analysis, is allowed to vary from one observation set to another.  The observational results of Vaillancourt \& Matthews (2012) of a variety of Galactic clouds and environments imply an approximate P$^{ga}_{\nu}$\% ratio of 10:17:20 for 350\micron:850\micron:1300\micron, with  P$^{ga}_{\nu}$\% ranging from approximately 5 to 10 at 350$\mu$m (Dotson et al., 2010), but we do not enforce this ratio or magnitude in our modelling.  Instead we look at each observation set independently and use a P$^{ga}_{\nu}$ value to produce the same mean polarization in a modelled observation set to the mean obtained in the real observation set.  We record the values of P$^{ga}_{\nu}$ used in Table 2.  

When calculating the Planck function B$_{\nu}$(T), we assume to first order a spatially constant dust temperature across the L1527 cloud core for the SHARP and SCUPOL observation sets,  based on the sub-mm color temperature observations of Chandler \& Richer (2000).  Chandler \& Richer argue that the large opening angle in the L1527 outflow provides the opportunity for heating away from the protostar through straylight from the protostar and possible shocks due to the outflow thus resulting in the near constant dust temperature observed on the scales of 10s of arcseconds.  

For the CARMA observation set, we include a temperature gradient for the inner central region using a temperature profile given in Kwon et al. (2009) of:
\begin{equation}
{T(r) = (T_o) (r/r_o)^{-2/(4+\beta)}}
\end{equation}
where $T_o = 12$ K, $r_o = 2000$ AU and $\beta= 0.5$ where $\beta$ is the dust opacity index.  These values for $T_o$,  $r_o$ and $\beta$ are consistent with the L1527 far-IR observations of Ladd et al. (1991) and the dust opacity modelling work of Kwon et al. (2009) for the inner regions of cloud cores containing Class 0 YSOs.  The low value of $\beta$ compared to the ISM value is due to grain growth. This assumed value of $\beta$ differs from the value of 1.8 used in column 6 of Table 1 in Ossenkopf \& Henning (1994) from which we obtained our value for  $\kappa_{\nu}$ for the CARMA observation set.  However, Ossenkopf \& Henning (1994) state in their conclusions that $\kappa_{\nu} = 1.11$ cm$^2$/g is still a reasonable value to adopt, even within the central dense inner regions of cores, but $\kappa_{\nu}$  may increase by a factor of 5 if the dust ice mantles have been removed from the dust aggregates. We will return to this possibility in our discussion of our results in Section 4.

The length of the line of sight used in our calculations of the model Stokes parameters in Appendix B (i.e.,  $\pm L_{\nu}(\alpha, \delta)$)  are wavelength dependent as well as sky position dependent.  $L_{\nu}(\alpha,\delta)$ has a wavelength dependence since we are using the wavelength to identify the observation set we are modelling.  CARMA observations, for example, sample much less volume of a YSO than do SCUPOL observations.  We assume the sample volume of a particular observation set to be nearly spherical with a radius of $R_{\nu}$ so the spatial dependence of $L_{\nu}(\alpha,\delta)$ is given by:
\begin{equation}
 {L_{\nu} (\alpha, \delta) = \sqrt{R_{\nu}^2 - d(\alpha, \delta)^2}, }
 \end{equation}
where  $d(\alpha, \delta) = \sqrt{(\alpha^2 + \delta^2)}D$, and D is the distance to L1527 (see Figure 5).  

Two processes restrict the effective extent of an observation, i.e., $R_{\nu}$: (1) the spatial filtering inherent in an observational technique; and (2) the sensitivity limits of a particular observation set.  A chop-nod technique was used for the SHARP and SCUPOL observations (Li et al. 2008; Greaves et al. 2003) with a 120$\arcsec$ chop-throw.  This means the flux exterior to a radius of 120$\arcsec$ is subtracted from the flux interior to this radius.  Thus, in effect, the extent of the observations are truncated in the plane of the sky (POS) and approximately along the line of sight (LOS) at a radius of this chop-throw.  In addition, the extent of the observations are also restricted by the sensitivity of the observations to the flux within the chop-throw region.   Observations of Chandler \& Richer (2000) illustrate that for SCUPOL this region is bounded by a radius of $\sim$90$\arcsec$, and for the wavelength used by SHARP the extent is less and is bounded by a radius of $\sim$60$\arcsec$.  The interferometric technique used by CARMA imposes a spatial filter in the POS of about 5 times the resolution, so on the order of 15$\arcsec$ for the 3$\arcsec$ resolution of the CARMA observations being used in this paper.  This same spatial filtering can be assumed along the LOS if one assumes that as one moves outward from the core into less dense material, it is likely that this lower density material has a greater angular extent than the core itself on the order of its distance from the core, and so is also spatially filtered in regions exterior to $\sim$ 5 times the resolution.  In addition,  the CARMA 3$\arcsec$ intensity map of L1527 in Hull et al. (2014) implies that most of the emission being measured by CARMA is well interior to a radius of 10$\arcsec$ from the YSO.  In summary, using the above, we have set the value of $R_{\nu}$ to be 12600 AU (i.e, 90$\arcsec$) for the SCUPOL data simulations,  8400 AU (i.e, 60$\arcsec$) for the SHARP data simulations, and 1400 AU (i.e, 10$\arcsec$) for the CARMA data simulations.

Once $I_{\nu}^{beam}$, $Q_{\nu}^{beam}$, and $U_{\nu}^{beam}$ are calculated using the equations in Appendix B for a given sky position, model, and observation set (defined by the subscript ``$\nu$''), the east from north orientation of the \underline{model} polarization-vector for that sky position can be given by:
\begin{equation}
{\xi_{\nu}(\alpha, \delta) = {1 \over 2} \arctan \biggl[{U_{\nu}^{beam} (\alpha, \delta) \over { Q_{\nu}^{beam} (\alpha, \delta)}} \biggr]}
\end{equation}
where the value of $\xi_{\nu}$ depends on which of one the four sign combinations exists for $U$ and $Q$, i.e.,:  $U \ge 0,\ Q \ge 0; \ \ U < 0, \ Q <  0; \ \ U < 0, \ Q > 0;$  or  $U > 0, \ Q < 0 $.

In addition, the percentage polarization fraction, P\%, of this vector can be given by:
\begin{equation}
{P\%_{\nu}(\alpha, \delta) = 100\biggl[{(U_{\nu}^{beam} (\alpha, \delta))^2 + (Q_{\nu}^{beam} (\alpha, \delta))^2 \over { (I_{\nu}^{beam} (\alpha, \delta)})^2}\biggr]^{1/2}.}
\end{equation}

The model polarization-vectors at each position on the sky can be rotated by 90$\arcdeg$ to produce ``B-vectors'' which represent the net orientation of the ``plane of the sky'' magnetic field after integration along a given line of sight.

\section{Results}

\subsection{An Overview of the Model B-Vector Results}

Figure 6 displays the modelled B-vectors for all four models (SA, WA, SO and WO) using the SCUPOL observation set parameters. For the two ``orthogonal'' models (SO and WO) two lines of sight are plotted separately (i.e., LOS-1 and LOS-2 as defined in Figure 4).  The SA and WA models are nearly symmetrical about their polar (rotation) axis, causing their B-vector maps in LOS-1 and LOS-2 to be almost identical.  In Figure 6, the polar (rotation) axis of the models has been rotated to lie east-west to be aligned with the rotation axis of L1527. The green arrows shown in Figure 6 represent the E-W outflow axis in L1527.

The B-vectors are plotted on a Nyquist-sampled spatial grid for a 20$\arcsec$ FWHM beam. The lengths of the vectors are proportional to the modelled polarization fraction at the location of the vector. The orientation of the vector gives the net orientation of the ``plane of the sky''  magnetic field at the sky position of the vector.  A P$_{\nu}^{ga}$ value of 10\% was used for all the models in Figure 6. The B-vectors are superimposed on the contour/greyscale intensity maps corresponding to the model indicated on the figure. The contours are in increments of one-tenth of the peak intensity.

Superimposed on each panel of Figure 6 are two areas ``A'' and ``B'' enclosed by green dot-dash lines. These delineate the areas within the B-vector maps that have the greatest differences from model to model.  It is noteworthy that all the models (for both LOS-1 and LOS-2) look similar in the four quadrants lying outside areas A and B. The B-vector orientations observed in these four quadrants are to be expected for the two ``aligned'' models (SA and WA), consistent with the classic hourglass magnetic field structure.  However, the same four quadrants in the SO and WO models produce similar inward pointing B-vectors.  Figure 4(b) illustrates why this is the case for both LOS-1 and LOS-2.  In a global cloud core collapse (as is the case in the models being studied in this paper), material from the outer part of the cloud core collapses onto the central region, dragging with it the field lines so they gain a large B component along the polar (rotation) axis.  This would not be as marked in an inside-out collapse, where the material collapsing to form the protostar was close to the center of the cloud core at the beginning of the collapse.

Figure 6 shows that it is the outer regions of area ``A'' that most clearly differentiate between an ``aligned'' model (e.g., SA and WA) and an ``orthogonal'' model (e.g., SO and WO).  And it is within area ``B'' that the differences in the spatial variation of the polarization fraction indicate differences in field strength.  The polarization fractions increase towards the center  within area ``B'' for the SA model, but decrease towards the center for the WA model. This is because the field lines become more twisted and toroidal within the central region of the core for the weaker aligned field, than is seen for the stronger aligned field.  The near toroidal nature of the central core B-fields in the WA model can be seen explicitly in the CARMA model observations, but not directly in the SCUPOL model observations (seen in Figure 6), except in the decrease in the polarization level towards the inner core as seen in panel (d) of Figure 6.

Figure 6(e) also shows a decrease in polarization fraction towards the center for the WO/LOS-1 model, in contrast to the large polarization fraction at the center of Figure 6(b) for the SO/LOS-1 model.  The reduction seen in Figure 6(e) in the outer regions of area ``B'' are due to field line distortions, but the reduction seen in the center of area ``B'' is dominated by the high optical depth in the center of the core for the WO model.  Recall, this is the only model of the four to form a circumstellar disk.  Without this optical depth effect, the polarization fraction would be high in the center of the WO model. 

Two cautionary notes in regards to the statements above for area ``B'':
(1) $P^{ga}_{\nu}$ is assumed spatially constant in our modelling, so the variations within area ``B'' are due only to variations in the magnetic field orientations along the line of sight and within the beam, and/or optical depth effects.  If, however, $P^{ga}_{\nu}$ varied spatially, then the variations in polarization strength and orientation in area ``B'' could be altered.  In Section 4.1.5 we argue, however, that $P^{ga}_{\nu}$ should be fairly constant within area ``B''. 
(2)  Area ``B'' is the region of the L1527 map where the bipolar outflow in L1527 could potentially alter the magnetic field structure from those of the models. More on this in Section 4.1.4.

Of all the models, the SO model has the largest variation between its two lines of sight (LOS-1 and LOS-2).  In the weaker field model, WO, the field lines have been distorted and twisted sufficiently to basically lose a distinction between LOS-1 and LOS-2.  However, the stronger fields of the SO model hold-on to this distinction; for the LOS-2 case, there are field lines with large polar-axis components superimposed on field lines along the same line of sight with large components orthogonal to the polar axis.  The result is an overall reduction in the effective polarization fractions for the LOS-2 case. The orientations of the B-vectors in the SO/LOS-2 model are more like those of the WO model than the SO/LOS-1 model.

\subsection{$\chi^2$ Comparisons between the L1527 Observations and the Models}

Figures 7, 8 and 9 compare directly each observed B-vector (red) to the model B-vectors (blue) for the SA, WA, SO/LOS-1, SO/LOS-2, WO/LOS-1 and WO/LOS-2 models.  Figure 7 covers this comparison for the SCUPOL observation set; Figures 8 and 9 for the SHARP and CARMA observation sets, respectively.  For both the observed and model B-vectors (inferred from the associated polarization vectors), the length of a vector is proportional to the polarization fraction of the polarization-vector associated with that B-vector. 

As stated before, we maintained the assumption that $P^{ga}_{\nu}$ is spatially constant in our modelling; therefore, any variation in the polarization levels across a map of a given observation set is due to variations in the magnetic field orientations along the line of sight, averaging over the beam size, and/or optical depth effects.  However, although spatially constant, we did allow $P^{ga}_{\nu}$ to vary in magnitude from one observation set to another and from one model to another so that:
\begin{equation}
{ \sum_{i=1}^{N_{obs}}P\%_{\nu}(\alpha_i, \delta_i) = \sum_{i=1}^{N_{obs}}P\%_{obs,i}}
\end{equation}
where $P\%_{obs,i}$ and $P\%_{\nu}(\alpha_i, \delta_i)$ are the actual and modelled percentage polarization fractions within an observation set (defined by $\nu$) at position $(\alpha_i, \delta_i)$, and $N_{obs}$ is the number of B-vectors detected in a given observation set.  The $P^{ga}_{\nu}$ values required to make Equation 12 hold are tabulated in Table 2. 

Figures 7, 8 and 9 illustrate that we must consider both the length of the B-vectors (i.e., the polarization fraction associated with each B-vector) as well as their orientation to assess the goodness-of-fit of a model to a given observation set.  We have calculated two $\chi^2$ statistics for each observation set, model and LOS-axis combination.  One $\chi^2$ statistic tests the goodness-of-fit in regards to the orientation of the B-vectors, $\chi_{ang}^2$; the other tests the goodness-of-fit in regards to the polarization fractions associated with the B-vectors,  $\chi_{p}^2$.

Our definition of $\chi_p^2$ is as follows:
\begin{equation}
{ \chi_p^2 = \sum_{i=1}^{N_{ind}} { [P\%_{obs,i} - P\%_{\nu}(\alpha_i, \delta_i)]^2\over {\sigma_{p,i}^2}} }
\end{equation}
where  $P\%_{obs,i}$ and $P\%_{\nu}(\alpha_i, \delta_i)$ are the observed and modelled polarization fractions of $N_{ind}$ independent polarization-vectors within a given observation set, and $\sigma_{p,i}$ are the 1-sigma uncertainties of the observed $P\%_{obs,i}$.  We can calculate the reduced-$\chi^2$ by dividing the above chi-square expression by the total number of the degrees of freedom, $(N_{ind}-1) $, where the reduction by 1 is because we have fitted one parameter ($P^{ga}_{\nu}$) via equation (12).  However, the spatial sampling of the data in each of the observation sets is less than a beam-width, so all the observed B-vectors in an observation set are not independent; there is some correlation between adjacent B-vectors.  Nevertheless, in our tabulation of our reduced-$\chi_p^2$ results in Table 2, we have assume that $N_{ind} = N_{obs}$ (i.e.,  the number of B-vectors detected in each observation set).  Since this does not represent the true number of independent terms in Equation (13), we will use this reduced-$\chi^2$ value only as an indicator of the goodness-of-fit; we will not use this statistic to evaluate any confidence levels in regards to this fit.

Similarly, our definition of $\chi_{ang}^2$ is:
\begin{equation}
{ \chi_{ang}^2 = \sum_{i=1}^{N_{ind}} { [\xi_{B,i} - \xi_{B,\nu}(\alpha_i, \delta_i)]^2\over {\sigma_{\xi,i}^2}} }
\end{equation}
where  $\xi_{B,i}$ and $\xi_{B,\nu}(\alpha_i, \delta_i)$ are the observed and modelled inferred position angles of $N_{ind}$ independent  B-vectors within a given observation set, and $\sigma_{\xi,i}$ is the 1-sigma uncertainty of the observed $\xi_{B,i}$.  Note: since the B-vectors are two headed vectors (i.e., there is a 180$^o$ degeneracy in their orientation), the angular difference in each term of Equation (14) must be $\le$ 90$^o$. If required, we added or subtracted 180$^o$ to make this so for each of the terms. We can calculate the reduced-$\chi_{ang}^2$ by dividing the above $\chi^2$ expression by the total number of the degrees of freedom, $N_{ind}$.  In this case we do not reduce this number by 1 since the model values of $\xi_{\nu}(\alpha_i, \delta_i)$ do not depend on the value of $P^{ga}_{\nu}$.  As with the reduced-$\chi_p^2$ case, we have tabulated our reduced-$\chi_{ang}^2$ values in Table 3 using $N_{ind} = N_{obs}$, and so we will use this value only to indicate the goodness-of-fit between a model and an observation set. 
  
In Table 3, we also list the values of the fraction $\chi_{ang}^2/\chi_{max}^2$ for each observation set, where:
\begin{equation}
{ \chi_{max}^2 = \sum_{i=1}^{N_{obs}} { [90]^2 \over {\sigma_{\xi,i}^2}} }
\end{equation}
is the maximum possible $\chi_{ang}^2$ value for an observation set.  If the value of this fraction is very close to zero, then the fit is good.  The opposite is true if this fraction is close to 1.

Finally we also include in Tables 2 and 3 the combined reduced-$\chi^2$ values of all three observation sets.  Combined $\chi_{ang}^2$
\begin{equation}
{ = \sum_{i=1}^9 { [\xi_{B,i} - \xi_{B,1300}(\alpha_i, \delta_i)]^2 \over {\sigma_{\xi,i}^2}} + \sum_{j=1}^{30} { [\xi_{B,j} - \xi_{B,350}(\alpha_j, \delta_j)]^2 \over {\sigma_{\xi,j}^2}} + \sum_{k=1}^{92} { [\xi_{B,k} - \xi_{B,850}(\alpha_k, \delta_k)]^2 \over {\sigma_{\xi,k}^2}} } 
\end{equation}
and this is divided by $N_{obs} = (9 + 30 + 92)$ to give the reduced chi-square statistic. Combined $\chi_p^2$
\begin{equation}
{= \sum_{i=1}^9 { [P\%_{obs,i} - P\%_{\nu}(\alpha_i, \delta_i)]^2\over {\sigma_{p,i}^2}} + \sum_{j=1}^{30} {  [P\%_{obs,j} - P\%_{\nu}(\alpha_j, \delta_j)]^2\over {\sigma_{p,j}^2}} + \sum_{k=1}^{92} { [P\%_{obs,k} - P\%_{\nu}(\alpha_k, \delta_k)]^2\over {\sigma_{p,k}^2}}} 
\end{equation}
and this is divided by $(N_{obs}-3) = (8 + 29 + 91)$ to give the reduced-$\chi^2$ statistic.

\section{Discussion of the Results}

\subsection{Implications of the Chi Squared Comparisons}
The results of our chi-squared comparisons between the models and the observations are presented in Tables 2 and 3. A good fit to the observations is indicated if $\chi_{ang}^2/N_{obs}$ and $ \chi_p^2 / (N_{obs}-1)$ are approximately equal to one, and if $\chi_{ang}^2/\chi_{max}^2$ is close to zero.  If the latter is close to one, this indicates a very poor fit to the data.

We list here three statements that can be made immediately from Tables 2 and 3 alone:\\
\noindent (i) Only for the SO model do the $\chi^2$ values change significantly when using the LOS-2 line of sight rather than the LOS-1 line of sight. Figure 6 illustrates why this would be the case. \\
(ii) Table 3 shows that the CARMA observation set rules out the SA model for L1527 - at least in the central region of L1527 - since the reduced chi-square values are very large and $\chi_{ang}^2/\chi_{max}^2$ is close to one. \\
(iii) The overall favored model is the WA model, but the SO/LOS-2 and WO models cannot be discarded.  No model fits the data within the errors of all the observation sets, but the WA, SO (for LOS-2 only) and WO models are not wildly at odds with the observations. 

Of the two orthogonal models, WO is favored over the SO model since the WO model fits the observations well in the center of L1527 (reduced-$\chi_{ang}^2 \sim 1$), and this model is not line of sight dependent as is the case for the SO model.  The SO/LOS-1 model is a poor fit to the SCUPOL observations. 

\subsubsection{ The Outer and Intermediate Regions of the L1527 Core: SCUPOL and SHARP}

Tables 2 and 3 imply the SHARP observation set is not a good differentiator between the models.  In part, this is because the SHARP vectors, in this particular observation set, happen to lie predominantly in a region outside Areas ``A'' and ``B'' (compare Figure 8 with Figure 6), in a region where the B-vectors are similar for all the models. But in addition, the SHARP $\chi^2_{ang}$ values for the WA, SO and WO models are very similar due to the dominance of the terms in the $\chi^2$-sum associated with the ill fitting B-vectors in the western portion of the sky maps shown in Figure 8 for these models.  If these vectors were ignored, the WO and SO/LOS-1 models would fit the SHARP data significantly better than the WA, SA and SO/LOS-2 models (see Figure 8).  

In contrast, the SCUPOL observation set has B-vectors inside Areas ``A'' and ``B'' (compare Figure 7 with Figure 6),making this observation set a good differentiator between the models.  Tables 2 and 3 show that the SCUPOL observation set favors the weak aligned (WA) model for the outer core of L1527. However, even for this model the fit is not within the errors of the observations.

Together the SCUPOL and SHARP observations imply a magnetic field structure that is like the WA model in the outer regions of the cloud core, but which shifts towards the slightly more toroidal WO model by the intermediate cloud core region.

\subsubsection{The Inner Region of the L1527 Core: CARMA}

Figure 9 shows that the B-vectors are predominantly toroidal for the WA, SO and WO models (i.e., orthogonal to the rotation axis of L1527) in the inner region of the L1527 cloud core.  Only the SA model has B-vectors still aligned with the rotation axis.  In the case of the WA model, poloidal fields initially aligned with the rotation axis have been twisted into this toroidal structure within the inner core of L1527. The weak orthogonal (WO)  model has a reduced-$\chi_{ang}^2$ value close to one within the inner core, which makes it the most favored within the central region of the L1527 core. However, the polarization fractions of the vectors of this model have a spatial distribution that differs significantly from that observed.  In addition, in order to achieve the polarization fractions observed, a high degree of dust grain alignment efficiency is required for this model (i.e., $P^{ga}_{\nu} \sim 30\%$).  This may actually be the case in this central region, where the radiative torques of the RAT alignment process could be very efficient (Lazarian, 2007).  However, another possibility is that the opacity, $\kappa_{\nu}$, for this inner region, which contains a circumstellar disk for this model, is much higher than 1.11 cm$^2$/g.  If a value of 5 cm$^2$/g is used for $\kappa_{\nu}$, as suggested as a possibility for circumstellar disks by Ossenkopf \& Henning (1994), then the WO model gives a better reduced-$\chi_p^2$ value of 2.51 for a $P^{ga}_{\nu} = 11.5\%$, with a reduced-$\chi_{ang}^2$ value of 1.10. Another possibility is that radiation escaping through the poles of the disk create a shallower temperature gradient within the central region of L1527, than the one given by Equation (8).  For a spatially constant temperature distribution for the inner region, the WO model has a reduced-$\chi_{ang}^2 = 1.03$ and a reduced-$\chi_p^2 = 2.37$ for $P^{ga}_{\nu} = 10.0\%$.   Interestingly, though, if these same variations in $\kappa_{\nu}$ or temperature gradient are applied to the WA model, then similar values for the WA model would result (i.e., reduced-$\chi_{ang}^2\sim1.0$, reduced-$\chi_p^2\sim 2.5 $ and $P^{ga}_{\nu}\sim 9\%$). 

 The above illustrates that if the inner central regions of the L1527 models (i.e., within a radius of $\sim$ 1$\arcsec$) are weighted less, either by blocking their radiation with higher opacity or reducing their temperature weighting, then both the WO and WA models have very similar polarizations.  This means that in the outer regions of the CARMA observations (i.e., beyond the central $\sim$ 1$\arcsec$ radius) the toroidal magnetic field structures seem similar for the two models. 

\subsubsection {Overall Model Review}

As stated previously, based on the magnetic field structure alone, the overall favored model is the WA model, although the WO model cannot be discarded especially for the intermediate and inner regions of the cloud core. No model gives a good fit within the errors of all the observations under our assumed model parameters.  One possible explanation for this is that the initial magnetic field configuration in L1527, before the onset of collapse, was neither aligned nor orthogonal.  To determine the most likely initial magnetic field orientation, $\psi_{_B}$, we modelled an uniform field and compared this to each observation set.  This model has one parameter, $\psi_{_B}$, which we varied for each observation set to obtain the best reduced-$\chi_{ang}^2$ value.   (Note: the reduced-$\chi_{ang}^2$ in this case has one fewer degrees of freedom since we are fitting the  $\psi_{_B}$ parameter to obtain the best fit.) 

Table 4 summarizes the results.  Although the reduced-$\chi_{ang}^2$ values reported in Table 4 for the comparison between the uniform fields and the observations are better than any reported in Table 3 for our collapse model comparisons with the observations, it is important to appreciate that a \underline{different} uniform field is required for each observation set.   The best-fit uniform field has an orientation of 67$\arcdeg$ east of north for the outer regions of the L1527 core (i.e., the SCUPOL observations). For the regions closer to the center of the L1527 core, the best-fit uniform field rotates towards the axis orthogonal to the rotation axis (i.e., 26$\arcdeg$ for the SHARP observations and 1$\arcdeg$ for the CARMA observations). It is interesting to note that these values for $\psi_{_B}$ correspond to the unweighted average of the orientations of the B-vectors,  $\xi^{ave}_{_B}$, in each observation set, if the averaging technique of Li et al. (2006) (and, independently, of Stephens et al., 2011) is used.  This averaging method uses normalized Stokes parameters $q_{_B}$ and $u_{_B}$, with polarization fractions set to one so:
\begin{equation}
{ \xi^{ave}_{_B} = {1 \over 2} \arctan \biggl[{u_{_B}^{sum} \over { q_{_B}^{sum}}} \biggr]}
\end{equation}
where
\[
{q^{sum}_{_B} = \sum_{i=1}^{N_{obs}} \cos(2 \xi_{_B,i}) \ \ \ and \ \ \ u^{sum}_{_B} = \sum_{i=1}^{N_{obs}} \sin(2 \xi_{_B,i}) }
\]
and where $\xi_{_B,i}$ is the orientation of the i-th B-vector within a given observation set.  (Note:  as in Eq (10), the value of  $\xi^{ave}_{_B}$ depends on which of one the four sign combinations exists for $u^{sum}_{_B}$ and $q^{sum}_{_B}$.) 

In Table 4, under the titles of the SA, WA, SO and WO models, we list the $\chi^2$ values for these models (for LOS-1 and LOS-2) in regards to the three uniform fields listed, with the B-vector in each observation set replaced by the respective \underline{model} B-vector for that observation set.  The $\chi^2$ values for the models in Table 4 show that both the aligned (SA and WA) models best fit the uniform field given for the outer regions of L1527 mapped by SCUPOL, but that the SA model is a very poor fit to the uniform field given for the inner regions mapped by CARMA.  This latter uniform field is best fit by the weak orthogonal (WO) model. Indeed, the WO model is also the best fit to the uniform field defined by the SHARP observation set for the intermediate regions of the L1527 cloud core. However, this model as well as the SO model give very poor fits ($\chi_{ang}^2/\chi_{max}^2 \sim 0.5$.) to a uniform field with an orientation derived from the average of the SCUPOL B-vectors (i.e., 67$\arcdeg$ east from north ) in the outer regions of the L1527 cloud core.  Table 4 emphazies once again, that the WA model is the only one of the four models that can to some degree accommodate all three uniform fields as they change from the outer regions of the L1527 cloud core to the inner region; although the WO model is favored when considering only the intermediate and inner regions of the core.

An initial uniform magnetic field tilted 67$\arcdeg$ east from north would not be too different from the large scale field traced in the less dense gas near the L1527 cloud core by the optical and near-IR polarization vectors shown in Figure 1 of Chapman et al. (2011).  These vectors imply an orientation of the field outside the L1527 cloud core of about 50$\arcdeg$ east of north, although this field need not maintain this orientation inside the denser gas of the Taurus filament containing L1527.  The values of 26$\arcdeg$ and 1$\arcdeg$ for the orientation of the uniform field models for the SHARP and CARMA observation sets, respectively, are consistent with the twisting of this initial field, orientated at 67$\arcdeg$ east of north, to a toroidal field about the rotation axis in the center of the L1527 core.  Further, a WA-type model may be more consistent with the SHARP and CARMA observation sets, if the initial field is oriented 67$\arcdeg$ east of north, rather than 90$\arcdeg$.

\subsubsection{The Bipolar Outflow in L1527}

Our model simulations do generate slow outflows, but these are not the bipolar CO outflows (driven by fast jets) that are observed in L1527 (and other Class 0 objects). These fast jets are probably powered by disk accretion close to the disk inner edge $\sim$0.1 AU or less from the protostar. Our core collapse simulations do not have the resolution down to such small scales (see Appendix A). Our models do incorporate, however, the bipolar cavities observed in L1527 that were carved out by the outflows (see Figure 5). The model magnetic field structures inside these cavities do not affect the model SCUPOL, SHARP and CARMA B-vectors, since there is no dust emission to polarize in these regions.  In order to test the sensitivity of our results to the presence of the outflow cavities, we also modelled the observed B-vectors that would result if the outflow cavities did not exist.  We found that our $\chi^2$ results listed in Tables 2 and 3 for the most part were unaltered down to the 10\% level.  This means for the most part the magnetic field structure inside the outflow cavities contributes relatively little to the integrated line of sight observations.  But is the outflow strong enough to alter the magnetic field structure adjacent to the outflow cavity walls to a significant depth so that the model observations would be significantly altered from the model observations used in this paper? The SA Model has the largest magnetic field strength near the surface of the outflow cavity walls for radii between 10$\arcsec$ and 60$\arcsec$ from the central protostar in L1527.  The field for this model is about 100 $\mu$G on the cavity surface at about 30$\arcsec$ from the protostar. This corresponds to a magnetic energy density of $\sim 4 \times 10^{-10}$ erg cm$^{-3}$, which is close to the thermal energy density of the gas at the same location on the outflow cavity wall.  Davidson et al.\ (2011) estimated the energy density of the bipolar outflow in L1527 to be $\sim 3 \times 10^{-7}$ erg cm$^{-3}$.  Thus the outflow has an energy density that is hundreds of times that of the magnetic field's energy density at the surface of the cavity. It is therefore likely that the outflow alters the magnetic field structure adjacent to the outflow cavities to some degree.  We may be witnessing this in the results shown in Figure 8 for the SHARP observation set. Indeed many of the SHARP B-vectors close to the outflow biconical cavities (in particular many of the ill fitting B-vectors in the western portion of the sky maps in Figure 8) could be influenced by the outflow, making the interpretation of the SHARP results somewhat complicated.

\subsubsection{The Spatially Uniform P$^{ga}$ Assumption}

In this paper, we have assumed spatially constant P$^{ga}$ values.  If the dust grains in a cloud core are aligned with radiative torques as predicted by Lazarian (2007), then P$^{ga}$ is enhanced for a dust grain that is illuminated by asymmetric radiation, where this radiation is at wavelengths less than the size of the dust grain.  This would mean that for starless cloud cores P$^{ga}$ would be large on the surface of the cloud core, but decrease to low values deep within the cloud core.  This decrease in polarization towards the center of starless cloud cores has been observed by Goodman et al. (1992, 1996),  Matthews et al. (2009), and more recently by Jones et al. (submitted).  This decrease can be countered to some degree by the radiation from an embedded source, such as a protostar, since such a source provides asymmetric radiation to the dust surrounding it. Evidence that this is the case is seen in Whittet et al.\ (2008), Vaillancourt and Matthews (2012), Zeng et al.\ (2013), and in the existence of polarization vectors measured by interferometers, such as CARMA, around deeply embedded YSOs.

There are a number of regions with asymmetric radiation in the L1527 cloud core: the outer areas of the cloud core, the areas near the embedded protostar, and the areas adjacent to the outflow cavities.  So although our assumption that P$^{ga}$ is spatially uniform throughout the L1527 cloud core may not be strictly true, it is probably not too inaccurate for L1527, especially throughout Area ``B'' (shown in Figure 6) where the dust is illuminated by amble asymmetrical radiation. 

It is not straightforward to predict the likely changes to our model B-vectors if P$^{ga}$ varies spatially within certain regions of L1527, but for our two main conclusions we can say:\\
\indent (i) If P$^{ga}$ increases towards the center in the CARMA observation set, then this would favor the WO model more than the WA model, based on our discussion in Section 4.1.2, so our overall conclusion in Section 4.1.2. would not change.\\
\indent (ii) If  P$^{ga}$ decreases from the outer edge of the L1527 cloud core, this would enhance the ability for the SCUPOL observation set to differentiate between an aligned model versus an orthogonal model since the outer regions of Area ``A'' would be enhanced. However, this recognized enhancement would not change our general results of Section 4.1.1 and 4.1.3 that favor the aligned models over the orthogonal models in the outer region of L1527 cloud core.  

\subsection {Comparisons to the Results of Others}

Observational studies of the initial magnetic field configurations in collapsing cloud cores and how these fields relate to the rotation axes of those cloud cores can be divided into two general categories: (1) studies with a statistical approach of a large number of sources (e.g. Hull et al., 2013; Krumholz et al., 2013; Hull et al., 2014); and (2) studies with a model approach comparing models to individual sources which have been selected for their relatively simple geometries and relative isolation on the sky (e.g. Davidson et al., 2011; Chapman et al., 2013; and Stephens et al., 2013).   The study outlined in this paper falls into the second category.

\subsubsection{Statistical Studies}

The results of Hull et al. (2013; 2014), a survey study with a sample of 30 star-forming cloud cores, imply that:\\
\indent (i) the orientation of the mean magnetic field in the outer regions for each cloud core is randomly distributed with respect to the bipolar outflow axis observed for that source; but \\
\indent(ii) the inner regions of the cores have two distinct distributions, with one set of sources having a random distribution of their mean magnetic field orientation with respect to their outflow axis, but the other set having a distribution that is preferentially orthogonal to their outflow axis.   This second distribution is marked by its low CARMA polarization fractions.

L1527 is in the Hull et al. (2014) sample and is a prototypical example of this latter low polarization set. Hull et al. used the published data of SCUPOL and SHARP (i.e., Matthews et al., 2009 and Davidson et al., 2011) to determine an average magnetic field orientation for the ``outer'' regions of this cloud core, weighted by the intensity of the continuum emission at the location of the B-vector and limited to vectors within a radius of 40$\arcsec$.  The value of this average is 38$\arcdeg$ east of north with a dispersion of $\pm$ 43$\arcdeg$, where the dispersion reflects the dispersion of the field line orientations within the source.   Hull et al. found the mean orientation of the magnetic field measured by CARMA for the inner region of L1527 is 3$\arcdeg$ east of north with a dispersion of $\pm$ 8$\arcdeg$. 

Our results above would suggest that the intensity weighted average for the core region within a radius of 40$\arcsec$ is not the best form of averaging to use when attempting to estimate the initial magnetic field orientation.  The intensity weighted average as well as the sample restriction of within a radius of 40$\arcsec$ favors the inner region of a cloud core, and so favors field orientations that may have been altered due to cloud core collapse.  Our unweighted average method (based on Li et al., 2006 and Stephens et al., 2011) of the B-field orientation in L1527 (i.e., 67$\arcdeg$ east of north for the SCUPOL observations) over the whole cloud core gives a better estimate of the initial orientation of the magnetic field in L1527. 

In addition, any averaging technique to determine the orientation of the initial cloud core magnetic field will potentially have large systematic errors associated with it if there are only a few B-vectors observed in the outer core regions. This potential is illustrated by Figure 6.  If a large number of B-vectors are measured throughout the outer cloud core, Figure 6 shows that an unweighted average of the orientations of these vectors would lead to an average value aligned with the rotation axis in the case of the SA and WA models, and orthogonal to the rotation axis for the WO and SO models.  However, if only a sparse number of B-vectors are observed in a cloud core, then Figure 6 illustrates how a variety of erroneous mean orientations could be estimated, especially for the orthogonal models.  In the case of L1527, a large number of B-vectors are observed in the outer regions of the cloud core, and so our estimate of the initial orientation probably does not suffer a large systematic error of this kind. 

The general results of Hull et al. (2014) are consistent with the global cloud core collapse models.  In the outer regions, the mean magnetic field orientation can take on a large range of values depending on the sky positions of the observed B-vectors, even if the underlying reason for the measured orientations is one particular model (e.g., WA Model or WO Model) from source to source.  

Not all the sources in the Hull et al. (2014) sample are like L1527; the sources with high polarization fractions in the inner region mapped by CARMA have random orientations of their magnetic vectors with respect to their outflows in the inner regions.  L1157, studied in detail by Stephens et al. (2013), is an example of a source with magnetic field vectors preferentially aligned with the outflow axis in the inner region mapped by CARMA.  

If the sources in the Hull et al. (2014) sample have one underlying initial magnetic field orientation with respect to their rotation axes, rather than a random distribution of initial orientations, then this initial magnetic field orientation must be more aligned than not to the rotation axis of the cloud, because it is only the aligned models (e.g., SA and WA) that can give a variety of field orientations within the central region of a collapsed cloud. The orthogonal models (e.g., SO and WO) always give orthogonal fields within the central region of a collapse.   An initial strong aligned field (like that in the SA model) would produce aligned fields with the outflow in the inner region mapped by CARMA, as observed in L1157.  Alternatively, an initial weak aligned field (like that in the WA model) would mimic the WO model in the inner region of a core as we have seen in this paper for L1527.  The other seemingly random orientations measured by CARMA could be due to field strengths that range between these two extremes.

The major problem with the WA model is that it does not produce the circumstellar, Keplerian disk observed in L1527.  But as stated before, this could be due to the ideal-MHD models being considered here, not including turbulence (Santos-Lima et al. 2012) and/or not including non-ideal MHD contributions (e.g., Ohmic dissipation; Dapp \& Basu 2010).  In addition, the models here have not considered initial fields that are oriented at angles other than 0 and 90$\arcdeg$.

\subsubsection{Model Studies}

Chapman et al. (2013) compared six star-forming cloud cores, selected because their bipolar outflows lie in the plane of the sky, thus implying their rotation axes also lie close to the plane of the sky.  Both L1527 and L1157 are included in this sample.  Chapman et al. (2013) compared the magnetic vectors measured for these six sources to the inside-out collapse model of Allen et al. (2003).  The initial magnetic field for this model is aligned with the rotation axis of the collapsing cloud core.  

 Myers et al. (1995) and Zhou et al. (1996) measured what could be interpreted as an inside-out collapse expansion wave in the core of L1527 at a radius of about 38$\arcsec$ ($\sim$ 5000 AU) from the core's center.  In an inside-out collapse, the magnetic field outside this radius should be little affected by the collapse, and so should display their initial uniform field configuration in projection.  This would mean that for the SO/LOS-1 and WO/LOS-1 models of the SCUPOL observation set (shown in Figures 7b and 7e), there should be a very low detection rate of B-vectors outside a radius of 38$\arcsec$ since the magnetic field lines would predominantly lie along the line of sight outside the expansion wave.  For the SO/LOS-2 and WO/LOS-2 models of the SCUPOL observation set (shown in  Figures 7c and 7f), the B-vectors outside this radius should be vertical, that is orthogonal to the rotation axis.  For lines of sight between LOS-1 and LOS-2, there would be only vertical vectors as well, with increasing length from LOS-1 to LOS-2.   SCUPOL data in Figure 1 rule out such an inside-out collapse for L1527 for an initial uniform field that is orthogonal to the rotation of the L1527 core.  The observations shown in Figure 2 are less problematic for an inside-out collapse if the initial uniform field is aligned (or nearly aligned) to the rotation of the core, and if some pre-collapse contraction has occurred, involving the whole core, prior to the collapse of the inner core.
 
Chapman et al. (2013) analysed only the SHARP data for their sources and so were looking at data mostly within the expansion wave radius.  In order to improve the signal-to-noise of the SHARP results and the number of vectors available to test the Allen et al. (2003) model, they combined the I, Q, and U data of the six sources after aligning and scaling them appropriately.  This approach implicitly assumes that all sources are undergoing inside-out core collapse.  The epochs of the collapse in each core were made the same by scaling their expansion envelopes so they coincided.  The sources were aligned via the axes of their observed flattened column density enhancements and outflows. The combined result is consistent with the Allen et al. (2003) model which has some pre-collapse core contraction (see Figure 8 of Chapman et al., 2013).  Interestingly, these combined results have a vector at the center of the core collapse more aligned than orthogonal to the outflow/rotation axis.  This is presumably because those sources in the sample of six that have fields orthogonal to the outflow/rotation axis in the center of the core have weak central polarizations, which would not contribute significantly to the integrations of Q and U in this location.  Very significant for this combined sample of 6 YSO cloud cores are the B-vectors nearly aligned with the rotation axis at a number of outer positions on the equatorial plane.  These vectors lie in the outer region of Area ``A'' in our Figure 6.  As mentioned earlier, this region has very different B-vector orientations, but similar polarization fractions, for the aligned and orthogonal models.   This combined result of Chapman et al. (2013) is more consistent with our SA and WA models than our SO and WO models, at least for these 6 sources, one of which is L1527.  

\section{Conclusions}

We have compared the polarization fractions and orientations of 92 B-vectors inferred from SCUPOL observations (Matthews et al., 2009), 30 B-vectors inferred from SHARP observations (Davidson et al., 2011; and this paper), and 9 B-vectors inferred from CARMA observations (Hull et al., 2014) with the corresponding model B-vectors of four ideal-MHD, non-turbulent global cloud core collapse models of the Class 0 YSO, L1527.  In general we conclude:

 (i) Very little of the initial cloud core magnetic field orientation exists after a global collapse of a rotating cloud core. Fields mapped on scales within a radius of 2000 AU (e.g., mapped by interferometers such as CARMA), in general can be significantly altered by the collapse.  Studies should not use inner core polarimetry results to estimate the initial orientation of the magnetic field in a cloud core before collapse.

(ii) The orientations of the B-vectors in the outer regions of the ``equatorial plane'', orthogonal to the rotation axis of a cloud core (i.e., outer regions of Area ``A'' shown in Figure 6), give the best estimate of the initial magnetic field orientation in that cloud core before collapse.    These regions should be the target of future polarimetry studies.  

 (iii) Another good estimate of the initial magnetic field orientation in a cloud core is an average using the unweighted summation of the normalized Stokes parameters, q and u, of each of the B-vectors in the cloud (as outlined in Section 4, based on Li et al., 2006 and Stephens et al., 2011).  However, this method will give accurate results only if a spatially well-sampled map of B-vectors is taken across the entire cloud core region, including its outer regions.  Figure 6 shows that in the case of a global cloud core collapse, there are many regions of the core that have very different B-vector orientations than the initial field orientation. 

 (iv)  In light of our global cloud core collapse models and averaging technique, a possible explanation to the survey results of Hull et al. (2014), could be that all the cloud cores in the Hull et al. (2014) sample survey had initial magnetic fields that were nearly aligned with the cloud core rotations (within $\pm 30\arcdeg$), and the differences seen between one core in the sample to the next can be explained by the difference in initial field strength in these sources rather than their initial orientations.  However, more thorough polarimetry studies of the outer regions of these cloud cores are needed in order to definitively determine which of the two explanations is correct.

(v) Our $\chi^2$ comparisons between the observed B-vectors and our model B-vectors in L1527 lead us to favor the weak aligned (WA) magnetic field model over the other three models since it is the only model that does not give bad fits to the B-vectors for all of the three L1527 observation sets.   No model with our assumed parameters fits all the observed B-vectors within the errors of the observations.  Although the WA model does not give a bad fit to the data in the inner region of L1527, the weak orthogonal (WO) model gives the best fit in this inner core. The WA model gives the best fit in the outer regions of the L1527 core.   Further, the $q$-$u$ unweighted average of the observed SCUPOL B-vectors implies a net B-vector orientation that is tilted with respect to the rotation axis by 23$\arcdeg$; this observed average orientation in the outer regions of L1527 is very unlikely for the WO model.  Our results are at odds with the predictions of the ideal-MHD global collapse models we are testing in this paper and the observational fact that there exists a circumstellar disk in L1527. The WA model cannot produce such a disk.  We therefore conclude that an ideal-MHD, non-turbulent global collapse model probably does not represent the core collapse that has occurred in L1527.

\section*{Acknowledgements}
J.A.D. is grateful to UWA for supporting her research through UWA Grant 10300026.    Z-Y.L. is supported in part by NASA NNX10AH30G, NNX14AB38G and NSF AST1313083.   C.L.H.H. and R.L.P. were supported in part by NSF grant AST1140031.   C.L.H.H. acknowledges support from an NSF Graduate Fellowship and from a Ford Foundation Dissertation Fellowship. LWL acknowledges NSF AST-1139950.  G.N. and N.L.C. would like to thank the National Science Foundation for their support of research with the SHARP polarimeter, via grant AST-0909030 to Northwestern University.  J.J.T. acknowledges support provided by NASA through Hubble Fellowship grant \#HST-HF-51300.01-A awarded by the Space Telescope Science Institute, which is operated by the Association of Universities for Research in Astronomy, Inc., for NASA, under contract NAS 5-26555. The National Radio Astronomy Observatory is a facility of the National Science Foundation operated under cooperative agreement by Associated Universities, Inc.   Support for CARMA construction was derived from the Gordon and Betty Moore Foundation, the Kenneth T. and Eileen L. Norris Foundation, the James S. McDonnell Foundation, the Associates of the California Institute of Technology, the University of Chicago, the states of Illinois, California, and Maryland, and the National Science Foundation. Ongoing CARMA development and operations are supported by the National Science Foundation under a cooperative agreement, and by the CARMA partner universities.

\appendix

\section*{APPENDIX A}

\subsection*{Technical Details of Model Simulations}

We start our simulations from a uniform, static, spherical core of $2 M_\odot$ and radius $2.25\times 10^{17}$~cm in a spherical coordinate  system $(r,\theta,\phi)$. We adopt an isothermal equation of state with a sound speed $a=0.2$~km~s$^{-1}$ below a critical density $\rho_c=10^{-13}$
g~cm$^{-3}$, and a polytropic equation of state $p\propto\rho^{5/3}$ above it. At the beginning of the simulation, we impose a solid-body
rotation on the core, with the rotation axis along the $\theta=0$ direction.
The initial magnetic field is uniform, tilting away from the rotation axis by an angle $\theta_0$. We consider two extreme cases with $\theta_0=0$ (aligned) and $90^\circ$ (perpendicular), respectively. As in Li et al. (2013), we choose a non-uniform grid of $120\times 64\times
60$. In the radial direction, the inner and outer boundaries are located at $r=10^{14}$ and $2.25 \times 10^{17}$ cm, respectively.
The radial cell size is smallest near the inner boundary ($5\times 10^{12}$~cm or $\sim 0.3$~AU). It increases outward by a constant
factor $\sim 1.07$ between adjacent cells. In the polar direction, we choose a relatively large cell size ($7.5^\circ$) near the polar axes,
to prevent the azimuthal cell size from becoming prohibitively small; it decreases smoothly to a minimum of $\sim 0.63^\circ$ near
the equator, where a rotationally supported disk may form. The grid is uniform in the azimuthal direction. The boundary
conditions in the azimuthal direction are periodic.  In the radial direction, we impose the standard outflow boundary conditions. Material leaving the inner radial boundary is collected as a point mass (the protostar) at the center. It acts on the matter in the computational domain through gravity. On the polar axes, the boundary condition is chosen to be reflective. Although this is not strictly valid, we expect its effect to be limited to a small region near the axis.

\section*{APPENDIX B}

\renewcommand{\theequation}{B-\arabic{equation}}

\subsection*{Modelling the Sub-mm Stokes Parameters observed for L1527}

Here we extend the solutions to the equations describing the transfer of linearly polarized light given in Jones (1989) by including dust emission.

If we consider a small section, $\Delta z$, along a given line of sight, the dust in that section will contribute to the observed polarization by (1) its own dust emission within $\Delta z$, and (2) the way it absorbs the radiation travelling towards the observer through $\Delta z$.  (Scattering is ignored for the wavelengths being modelled in this paper.)

Jones (1989) derives solutions to the equations formulated by Jones and O'Dell (1977) describing the transfer of linearly polarized light through a non-birefringent medium with no internal emission.  These are the equations governing contribution (2) above.  Jones (1989) assumes that within $\Delta z$ the grains are partially aligned in one direction given by a position angle.  In this paper, we will use the position angle for the \underline{maximum} extinction within $\Delta z$ (i.e., the position angle of the \underline{long} axis of the dust grains within $\Delta z$ as projected onto the sky); this is the east from north angle $\xi$ defined in Section 2.3 of this paper.  

Based on the results of Jones (1989), the Stokes parameters of the exit radiation from $\Delta z$ (ignoring dust emission within $\Delta z$) are:
 
 \[
 {I = (I_o \cosh \Delta \tau_p + \zeta \sinh \Delta \tau_p ) e^{-\Delta \tau} }
 \]
 \begin{equation}
 {Q = (Q_o - I_{p} \cos 2\xi)e^{-\Delta \tau} }
 \end{equation}
 \[
 {U = (U_o - I_{p} \sin 2\xi)e^{-\Delta \tau} }
 \]
  where
\[
{\zeta = - Q_0\cos2\xi - U_o \sin2\xi \ ; \ \ \ \ I_{p} = I_o\sinh \Delta\tau_p + \zeta \cosh \Delta\tau_p - \zeta \ ;}
\]
 $I_o$, $Q_o$, and $U_o$ are the Stokes parameters of the incident radiation from behind, and 
\begin{equation}
{\Delta \tau =  {\Delta \tau_{\parallel} + \Delta \tau_{\perp} \over 2} = {\kappa \over 100}\rho_{_{gas}}(z_o) \Delta z \ ;  \ \ \ \ \  \Delta \tau_p = {\Delta \tau_{\parallel} - \Delta \tau_{\perp} \over 2}=P_o \Delta \tau .}
\end{equation}
$\Delta \tau_{\parallel}$ and $\Delta \tau_{\perp}$ are the optical depths for the radiation $\bold{E}$-vectors aligned and orthogonal to the long axis of the dust grains within $\Delta z$, respectively. $P_o(z)$ is the polarization fraction within $\Delta z$ due to absorption of unpolarized light in the optically thin limit. This can be further factorized to $P_o = P^{ga} \cos^2\gamma(z)$, where $P^{ga}$ depends on the dust grains and the degree of alignment of their rotation axes within $\Delta z$; and $\cos^2\gamma$ is the decrease in this polarization, as viewed by an observer, when the average rotation axes of the dust grains is inclined from the plane of the sky by $\gamma$.   

As stated before, Jones (1989) does not include the dust emission within $\Delta z$, we will do this here.  Within $\Delta z$, the dust emission $I_{em}$ and polarized emission $I_{P_{em}}$ for the \underline{non-optically thin} case with significant grain alignment can be given by:
\[
{I_{em}= I_{\parallel em} + I_{\perp em} = \int_0^{\Delta \tau_{\parallel} } B_{\nu}(T) e^{-\tau_{\parallel}} {d \tau_{\parallel} \over 2} + \int_0^{\Delta \tau_{\perp}} B_{\nu}(T) e^{-\tau_{\perp}} {d \tau_{\perp} \over 2} }
\]
\begin{equation}
{ = \biggl[1 - \biggl(\cosh(\Delta \tau_p)\biggr) e^{- \Delta \tau} \biggr] B_{\nu}(T),}
\end{equation}
where $B_{\nu}(T)$ is the Planck function at the frequency (wavelength) of interest, and
\[
{I_{P{em}}= I_{\parallel em} - I_{\perp em} = \int_0^{\Delta \tau_{\parallel} } B_{\nu}(T) e^{-\tau_{\parallel}} {d \tau_{\parallel} \over 2} - \int_0^{\Delta \tau_{\perp}} B(T)_{\nu} e^{-\tau_{\perp}} {d \tau_{\perp} \over 2} }
\]
\begin{equation}
{ =  \biggl(\sinh(\Delta \tau_p)\biggr) e^{- \Delta \tau} B_{\nu}(T)}
\end{equation}
(see Dennison, 1977). The $Q_{em}$ and $U_{em}$ Stokes parameters for this emission would then be given by $I_{P{em}}\cos 2\xi$ and $I_{P{em}}\sin 2\xi$, respectively.  It is important to note here that the only assumption for these equations to hold is that the alignment is constant within $\Delta z$ (i.e., both $P_o$ and $\xi$ are constant within $\Delta z$); both $\Delta \tau$ and $\Delta \tau_p$ can be very much greater than one for $\Delta z$.

Combining (B-1), (B-3) and (B-4), we cover both polarizing contributions made by the dust within $\Delta z$, i.e., by absorption and emission.  However, this is for just one section of a given line of sight. We must stepwise combine these contributions all the way along a given line of sight, section by section, starting at the rear of the cloud, defined by $L_{\nu}(\alpha, \delta)$, and moving forwards to the front of the cloud to $-L_{\nu}(\alpha, \delta)$ as defined in Figure 5.  The ``exit'' Stokes parameters for each previous section becomes the ``incident'' Stoke parameters for the next section.  The observed Stokes parameters are those that exit the last section in a given line of sight at $-L_{\nu}(\alpha, \delta)$.  

In other words, if there are N steps (1 through N) from the back of the cloud (at $L_{\nu}(\alpha, \delta)$) to the front of the cloud (at $-L_{\nu}(\alpha, \delta)$ for a line of sight at the sky position $(\alpha, \delta)$, then the step-by-step sequence followed to produce the observed Stokes parameters exiting the ``Nth'' step can be summarized as follows:

 \begin{equation}
 {I_{i+1} = (I_i \cosh \Delta \tau_{p,i} + \zeta_i \sinh \Delta \tau_{p,i} ) e^{-\Delta \tau_i} + (1-(\cosh \Delta \tau_{p,i})e^{-\Delta \tau_i}) B_{\nu}(T_i)}
 \end{equation}
  \begin{equation}
 {Q_{i+1} = (Q_i - I_{p,i} \cos 2\xi_i)e^{-\Delta \tau_i}  + (\sinh\Delta \tau_{p,i}) e^{-\Delta \tau} B_{\nu}(T_i) \cos 2 \xi_i}
 \end{equation}
  \begin{equation}
 {U_{i+1} = (U_i - I_{p,i} \sin 2\xi_i)e^{-\Delta \tau_i}  + (\sinh\Delta \tau_{p,i}) e^{-\Delta \tau} B_{\nu}(T_i) \sin 2 \xi_i}
 \end{equation}
 where 
 \[
{\zeta_i = - Q_i\cos2\xi_i - U_i \sin2\xi_i \ ; \ \ \ \ I_{p,i} = I_i\sinh \Delta\tau_{p,i} + \zeta_i \cosh \Delta\tau_{p,i} - \zeta_i \ ; \ \ \ \ \xi_i = \xi(\alpha, \delta, z_i) \ ; }
\]
\begin{equation}
{\Delta \tau_i = {\kappa_{\nu} \over 100}\rho_{_{gas}}(\alpha, \delta, z_i) \Delta z \ ;  \ \ \ \ \  \Delta \tau_{p,i} = P_{\nu}^{ga} \cos^2\gamma(\alpha, \delta, z_i) \Delta \tau_i }
\end{equation}
and where $I_o = 0$, $Q_o = 0$, and $U_o = 0$ for the first step, $i=1$ (i.e., assuming no emission outside the cloud). 

The quantities $\rho_{_{gas}}$, $\gamma$, and $\xi$ are determined by our ideal-MHD global collapse models for L1527.  The quantities $\kappa_{\nu}$,  $P_{\nu}^{ga}$, $T_i = T(\alpha, \delta, z_i)$, and $L_{\nu}(\alpha, \delta)$ all have assumed values for each observation set being modelled.  These assumptions are discussed in Section 2.3 of the paper.

In this formulation, the model Stokes parameters on the sky are:
\begin{equation}
{ I_{\nu}(\alpha, \delta) = I_{N+1}; \ \ \   Q_{\nu}(\alpha, \delta) = Q_{N+1}; \ \ \  and  \ \ \ U_{\nu}(\alpha, \delta) = U_{N+1}}
\end{equation}
for the line of sight at $(\alpha, \delta)$ for each observation set (indicated by the subscript ``$\nu$'') and for each cloud collapse model.

Before the above $I_{\nu}$, $Q_{\nu}$ and $U_{\nu}$ values can be compared with observations, they must be convolved with an appropriate Gaussian beam.  For a Gaussian beam,  $G_{\nu}$, centred on the sky position $(\alpha_o, \delta_o)$, the beam convolved measurements are:
\begin{equation}
{I_{\nu}^{beam}(\alpha_o, \delta_o) = \int_{\alpha} \int_{\delta} G_{\nu}\biggl(\alpha_o - \alpha, \delta_o - \delta \biggr) I_{\nu}(\alpha, \delta) d\alpha d\delta}
\end{equation}
\begin{equation}
{Q_{\nu}^{beam}(\alpha_o, \delta_o) =  \int_{\alpha} \int_{\delta} G_{\nu}\biggl(\alpha_o - \alpha, \delta_o - \delta \biggr) Q_{\nu}(\alpha, \delta) d\alpha d\delta}
\end{equation}
\begin{equation}
{U_{\nu}^{beam}(\alpha_o, \delta_o) =  \int_{\alpha} \int_{\delta} G_{\nu}\biggl(\alpha_o - \alpha, \delta_o - \delta \biggr) U_{\nu}(\alpha, \delta) d\alpha d\delta}
\end{equation}

 For our calculations we used 2-D circular Gaussian beams with the following form:
\begin{equation}
{ G_{\nu}(\Delta \alpha, \Delta \delta) =  \exp- \biggl[ {\Delta \alpha^2 + \Delta \delta^2 \over {\Omega_{\nu}^{^2}}}\biggr]}
\end{equation}
where 
\[
{\Omega_{\nu} = {FWHM \over {2 \sqrt{ \ln 2}}}}
\]
and where FWHM is the full-width half-maximum of the Gaussian beam being considered.

\begin{deluxetable}{cccccc}
\tablewidth{0pt}
\footnotesize
\tablecaption{Re-Reduced SHARP Polarimetry Results\label{tbl-specs}}
\tablehead{ \colhead{$\Delta \alpha$\tablenotemark{a}} & \colhead {$\Delta \delta$\tablenotemark{a}} & \colhead{$P$} & \colhead{$\sigma_p$} & \colhead{$\phi$\tablenotemark{b}} & \colhead{$\sigma_{\phi}$} \\
 \colhead{(arcsec)} & \colhead{(arcsec)}   &  \colhead{(\%)} & \colhead{(\%)} & \colhead{(deg)} & \colhead{(deg)} }
\startdata
 -33.3 & -3.4 & 9.1 & 4.4 & 63.9 & 11.8\\
 -33.3 & -8.2 & 11.5 & 4.8 & 57.7 &10.3\\
 -28.5 & 10.8 &  6.6 & 3 & -9 & 10.3\\
 -23.7 & 10.8 & 5.1 & 2  & -3 & 9.6\\
-19.0 & 10.8 & 3.5 & 1.5 &10.8 & 11\\
 -9.5 & -8.2 & 2 & 0.9 & -5 &11.4\\
 -9.5 & -22.4 & 4.6 & 2.3 & 41 &12.4\\
 -4.7 & -22.4 & 4.2 &1.9 & 30.4 &11.8\\
14.3 & 10.8 & 2.2 & 1& -51.9 & 12.3\\
 14.3 & 15.5 & 3.2 & 1.2 & -55	& 10.1\\
 14.3 & 20.3 & 4 & 1.6 &-50.8	&10.4\\
 19.0 & -17.7 & 2.4 & 1& 76.8 & 11\\
 19.0 & -12.9 &1.9& 0.9 &67.9 &11.9\\
 19.0 & 6.0 & 2.6	& 1.1 &	-78 &10.5\\
 19.0 & 10.8 & 2.7 &  1.2 &  -64.8 &	11.3\\
 19.0 &	15.5	 & 4.1 & 	1.3  & -54.9 &	8.9\\
 19.0 &	20.3	&  5.8  &	1.8	&  -52.2 & 8.3\\
 19.0 &	25.1	& 6.5 &	2.7 & -50.9 & 	10.3\\
 19.0 &	29.8	& 7.2 &	3.6 & -56.6 &	12\\
 23.8 &	6.0	& 3.1 &	1.2 & -81.1 & 10.5\\
 23.8 &	10.8 & 3 &	 1.4	& -68.1 &	11.9\\
 23.8 &	15.5	& 4.9 &	1.6	& -58.8 &	8.8\\
 23.8 &	20.3	& 5.3 &	2.1	& -59.9 &	10.3\\
 28.5 &	10.8 & 4.8	& 1.8 &	-74.4 &	9.8\\
 28.5 &	15.5 &	6.8	& 2	& -67.8 &	8.1\\
 28.5 &	20.3 &	5.3 &	2.4 &	-69 &	11.7\\
 33.3 &	10.8	& 6	& 2.3 &	-68.2 &	10.3\\
 33.3 & 15.5 & 7.1	& 2.5 &	-65.2 &	9.4\\
 38.0 &	10.8 & 6.6	& 3.2 & -58.6 & 12.3\\
 38.0 &	15.5 & 8 & 3.8 & -52.9 & 11.5\\
\hline
\enddata
\tablenotetext{a}{Offsets from YSO position:  04$^h$:39$^m$:53.$^s$9 \ 26$\arcdeg$:03$\arcmin$:09.$\arcsec$6  (J2000) 
 (based on CARMA measurements, Hull et al., 2014) }
\tablenotetext{b}{Position Angle of the polarization E-vector, measured east of north}
\end{deluxetable}

 \begin{landscape}
\begin{center}
\begin{deluxetable}{lccccccccccc}
\tabletypesize{\footnotesize}
\tablewidth{8.5in}
\tablecaption{ $\chi^2$ Comparisons between the Polarization Distributions of the Models and the Observations \tablenotemark{a} \label{tbl-specs}}
\tablehead{
\colhead{}    &  \multicolumn{2}{c}{SA Model} & &  \multicolumn{2}{c}{WA Model} & &  \multicolumn{2}{c}{SO Model} & & \multicolumn{2}{c}{WO Model} \\ 
\colhead{Observation Set } \hspace{1.1in}  &  \colhead{ $ {\chi_{p}^2 \over {N_{obs}-1}}$} & \colhead {P$_{\nu}^{ga}$} &  \hspace{0.2in} & \colhead{ $ {\chi_{p}^2 \over {N_{obs}-1}}$} & \colhead {P$_{\nu}^{ga}$} & \hspace{0.2in} &  \colhead{ $ {\chi_{p}^2 \over {N_{obs}-1}}$} & \colhead {P$_{\nu}^{ga}$}  & \hspace{0.2in} &  \colhead{ $ {\chi_{p}^2 \over {N_{obs}-1}}$} & \colhead {P$_{\nu}^{ga}$} }
\startdata
CARMA(N$_{obs}$= 9)\tablenotemark{b}  LOS-1 & 1.25  & 8.3\% &   & 4.75 & 14.3\% & & 3.88 & 23.4\%& & 7.63& 31.5\%\\
CARMA(N$_{obs}$ = 9) LOS-2 & 1.38  & 8.1\% &  & 6.00  &  14.6\% & & 3.88 & 11.0\% & & 7.25 & 28.4\% \\
SHARP(N$_{obs}$ = 30)  LOS-1 & 1.90  & 8.8\% &  & 0.79   &  15.3\% & & 1.90 & 6.3\% & & 2.28 &  11.8\%\\
SHARP(N$_{obs}$ = 30)  LOS-2 & 1.79  & 8.6\%  & & 0.79   &  15.3\% & & 1.10 & 16.6\% & & 2.07 & 16.5\% \\
SCUPOL (N$_{obs}$ = 92) LOS-1 & 21.5  & 12.5\%  & & 3.10  &  12.2\% & & 18.5& 12.0\% & & 8.56 &  16.6\%\\
SCUPOL (N$_{obs}$ = 92) LOS-2 & 20.8  & 12.5\% &  & 4.42  &  12.0\% & & 4.01 & 26.3\% & & 6.08 &  17.2\%\\
\hline
Combined (N$_{obs}$ = 131) LOS-1 & 15.8\tablenotemark{c}  &  &  & 2.68\tablenotemark{c}  &   & & 13.8\tablenotemark{b} &  & & 7.08\tablenotemark{c} &  \\
Combined (N$_{obs}$ = 131) LOS-2 & 15.3\tablenotemark{c}  &  &  & 3.70\tablenotemark{c} &  & & 3.34\tablenotemark{b} &  & & 5.24\tablenotemark{c} &  \\
\enddata
\tablenotetext{a}{\small{The first column for each model is the reduced-$\chi^2$ comparison between the polarization fraction of the polarization vectors observed to those in the model. The second column for each model shows the model P$_{\nu}^{ga}$ required to give the same average polarization value over all the modelled vectors in an Observation Set as those observed. These would be the polarization fractions observed if the magnetic fields throughout the cloud core were oriented in the same direction and lying on the plane of the sky. The models are defined in Section 3.}}
\tablenotetext{b}{\small{N$_{obs}$ is the number of polarization vectors observed on the sky for each Observation Set.}}
\tablenotetext{c}{\small{Derived using the formula $ \chi_{p}^2 / (N_{obs}-3)$}}
 \end{deluxetable}
 \end{center}
 \end{landscape}

\begin{landscape}
\begin{center}
\begin{deluxetable}{lccccccccccc}
\tabletypesize{\footnotesize}
\tablewidth{8.5in}
\tablecaption{ $\chi^2$ Comparisons between the B-Vector Angles of the Models and the Observations \tablenotemark{a}  \label{tbl-specs}}
\tablehead{
\colhead{}    &  \multicolumn{2}{c}{SA Model} & &  \multicolumn{2}{c}{WA Model} & &  \multicolumn{2}{c}{SO Model} & & \multicolumn{2}{c}{WO Model} \\ 
\colhead{Observation Set } \hspace{1.4in}  &  \colhead{ $ {\chi_{ang}^2 \over {N_{obs}}}$} & \colhead {$ {\chi_{ang}^2 \over {\chi_{max}^2}}$} &  \hspace{0.2in} & \colhead{ $ {\chi_{ang}^2 \over {N_{obs}}}$} & \colhead {$ {\chi_{ang}^2 \over {\chi_{max}^2}}$} & \hspace{0.2in} & \colhead{ $ {\chi_{ang}^2 \over {N_{obs}}}$} & \colhead {$ {\chi_{ang}^2 \over {\chi_{max}^2}}$} & \hspace{0.2in} & \colhead{ $ {\chi_{ang}^2 \over {N_{obs}}}$} & \colhead {$ {\chi_{ang}^2 \over {\chi_{max}^2}}$} }
\startdata
CARMA(N$_{obs}$ = 9)\tablenotemark{b}  LOS-1 & 248   & 0.919 &   & 9.89 & 0.037 & & 12.7 & 0.047& & 1.53& 0.006\\
CARMA(N$_{obs}$ = 9) LOS-2 & 237  & 0.880 &  & 7.56  &  0.028 & & 4.62 & 0.017& & 1.61 & 0.006 \\
SHARP(N$_{obs}$ = 30)  LOS-1 & 19.7   & 0.266 &  & 12.7   &  0.171 & & 9.50 & 0.128 & & 11.5 &   0.155\\
SHARP(N$_{obs}$ = 30)  LOS-2 & 19.7  & 0.266  & & 12.6   &  0.171 & & 16.4 & 0.222 & & 11.9 & 0.161 \\
SCUPOL (N$_{obs}$ = 92) LOS-1 & 24.7  & 0.184  & & 24.5  &  0.183 & & 45.1& 0.336 & & 38.7 &  0.288\\
SCUPOL (N$_{obs}$ = 92) LOS-2 & 24.8  & 0.185 &  & 24.4  &  0.182 & & 31.3 & 0.234 & & 40.0 &  0.298\\
\hline
Combined (N$_{obs}$ = 131) LOS-1 & 38.9  & 0.300 &  & 20.8  &  0.161 & & 34.7 & 0.268 & & 29.9 &  0.231\\
Combined (N$_{obs}$ = 131) LOS-2 & 38.2  & 0.295 &  & 20.5  &  0.158 & & 26.1 & 0.201 & & 30.9 &  0.238\\
\enddata
\tablenotetext{a}{\small{The first column for each model is the reduced-$\chi^2$ comparison between the orientation of the B-vectors observed to those in the model. The second column for each model shows how this $\chi^2$ compares to the maximum $\chi^2$ value for the case when the angular difference of all B-vectors between the model and the observations are 90$\arcdeg$ with respect to each other.  This second column value for each model will lie between 0 and 1; for a good fit this value is close to 0, for a bad fit  it is close to 1. The models are defined in Section 3.}}
\tablenotetext{b}{\small{{N$_{obs}$ is the number of B-vectors observed on the sky for each Observation Set.}}}
 \end{deluxetable}
 \end{center}
 \end{landscape}

 \begin{landscape}
\begin{center}
\begin{deluxetable}{lccccccccccccccc}
\tabletypesize{\footnotesize}
\tablewidth{8.5in}
\tablecaption{ $\chi^2$ Comparisons between the Observed and Model B-Vectors and the Best-Fit Uniform Field \label{tbl-specs}}
\tablehead{
\colhead{}    & &  \multicolumn{2}{c}{Observations} &  \multicolumn{2}{c}{SA Model} &   \multicolumn{2}{c}{WA Model} &   \multicolumn{2}{c}{SO Model} & \multicolumn{2}{c}{WO Model} \\ 
\colhead{Observation Set }  &  \colhead{\ \ \ \ $\psi_{_B}$\tablenotemark{a}}  &  \colhead{ $ {\chi_{ang}^2 \over {(N_{obs}-1)}}$} & \colhead {$ {\chi_{ang}^2 \over {\chi_{max}^2}}$} & \colhead{ $ {\chi_{ang}^2 \over {(N_{obs}-1)}}$} & \colhead {$ {\chi_{ang}^2 \over {\chi_{max}^2}}$} &  \colhead{ $ {\chi_{ang}^2 \over {(N_{obs}-1)}}$} & \colhead {$ {\chi_{ang}^2 \over {\chi_{max}^2}}$} &  \colhead{ $ {\chi_{ang}^2 \over {(N_{obs}-1)}}$} & \colhead {$ {\chi_{ang}^2 \over {\chi_{max}^2}}$} & \colhead{ $ {\chi_{ang}^2 \over {(N_{obs}-1)}}$} & \colhead {$ {\chi_{ang}^2 \over {\chi_{max}^2}}$} }
\startdata
CARMA(N$_{obs}$ = 9)\tablenotemark{b}    & 1 & 0.97 & 0.003 & 278 & 0.917 & 12.2& 0.040   &13.0  & 0.043  & 0.63 & 0.002 \\
 \ \ \  (LOS-2) &  &  &   &  271 & 0.895  & 7.07 & 0.023   & 3.31 & 0.011  & 0.68 &0.002 \\
SHARP(N$_{obs}$ = 30)      & 26 & 11.3 & 0.148 & 27.5& 0.359 & 16.1& 0.211 & 4.55 & 0.059  & 6.17 & 0.081 \\
 \ \ \ (LOS-2) & &  &   & 27.4 &  0.358 &16.1& 0.211& 21.6  & 0.283 & 9.23 & 0.121  \\
SCUPOL (N$_{obs}$ = 92)   & 67 & 21.2 & 0.157  & 18.4 & 0.135  & 20.4& 0.150 & 71.6 & 0.527  & 61.6 & 0.454 \\
 \ \ \ (LOS-2) &  &  &   &  18.1& 0.133 & 20.1 & 0.148  & 51.3 & 0.378 &  63.9 & 0.471 \\
\enddata
\tablenotetext{a}{\small{$\psi_{_B}$ is the east of north angle of the uniform B-field that best fits the data for each Observation Set.}}
\tablenotetext{b}{\small{N$_{obs}$ is the number of B-vectors observed on the sky for each Observation Set.}}
\end{deluxetable}
 \end{center}
 \end{landscape}

\newpage

\begin{figure}[p]
\begin{center}
\includegraphics[width=4.5in]{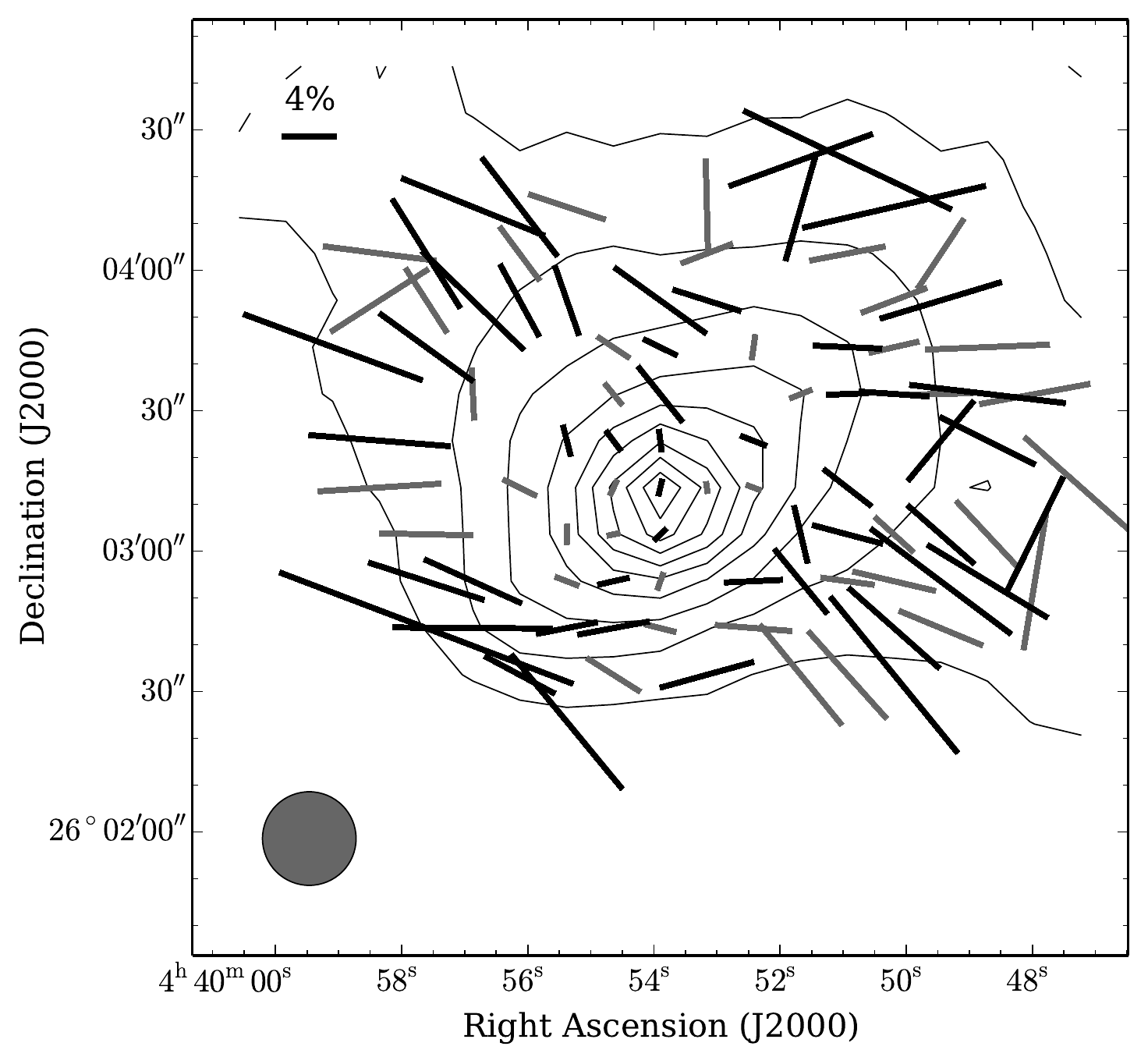}
\caption{\small{
The B-vectors inferred from the SCUPOL 850 $\mu$m polarimetry data of Matthews et al. (2009).  The length of each vector is proportional to the polarization fraction, P. The black vectors have $P/\sigma_{_P} \ge  3$, while the grey vectors have $2 \le P/\sigma_{_P} <  3$. The vectors are superimposed on the 850 $\mu$m continuum contour map measured by SCUPOL with an effective resolution of 20$\arcsec$. These contours are linear, in increments of one-tenth of the peak. 
}}  
\end{center}
\label{fig:IC348b}
\end{figure}

\begin{figure}[p]
\begin{center}
\includegraphics[width=4.5in]{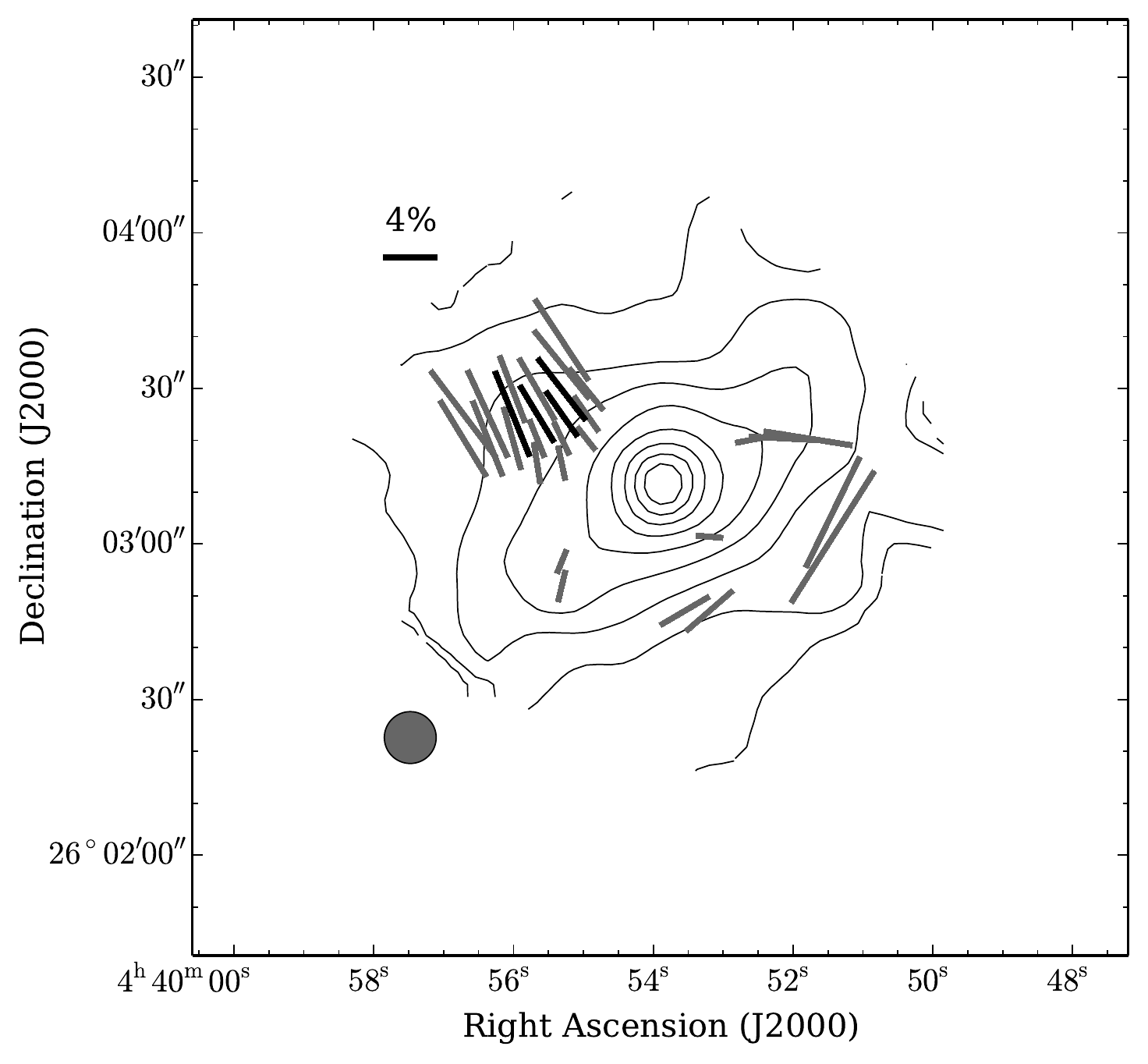}
\caption{\small{
The B-vectors inferred from the SHARP 350 $\mu$m  polarimetry data of Davidson et al. (2011), which have been re-binned and re-reduced here in a Nyquist sample of spatial vectors instead of a beam sample.  The length of each vector is proportional to the polarization fraction, P.  The black vectors have $P/\sigma_{_P} \ge  3$, while the grey vectors have $2 \le P/\sigma_{_P} <  3$. The vectors are superimposed on the 350 $\mu$m continuum contour map measured by SHARP with an effective resolution of 10$\arcsec$. These contours are linear, in increments of one-tenth of the peak. 
}}  
\end{center}
\end{figure}

\begin{figure}[p]
\begin{center}
\includegraphics[width=4.5in]{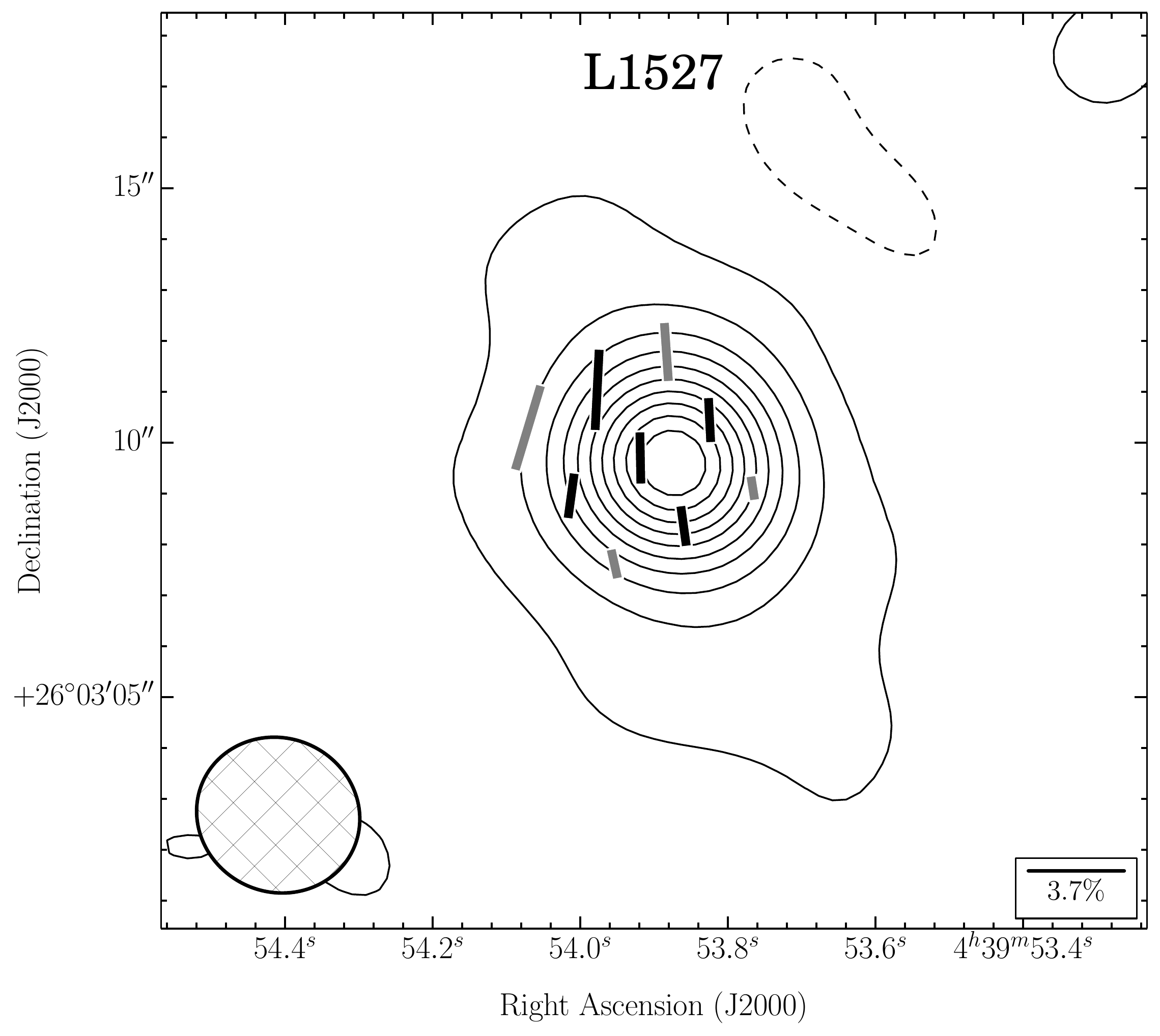}
\caption{\small{
The B-vectors inferred from the CARMA 1300 $\mu$m polarimetry data of Hull et al. (2014).  The length of each vector is proportional to the polarization fraction, P. The black vectors have $P/\sigma_{_P} \ge  3$, while the grey vectors have $2 \le P/\sigma_{_P} <  3$. The vectors are superimposed on the 1300 $\mu$m continuum contour map measured by CARMA with an effective resolution of 3$\arcsec$. These contours are linear, in increments of one-tenth of the peak. 
}}  
\end{center}
\end{figure}

\begin{figure}[p]
\begin{center}
\includegraphics[width=6.5in]{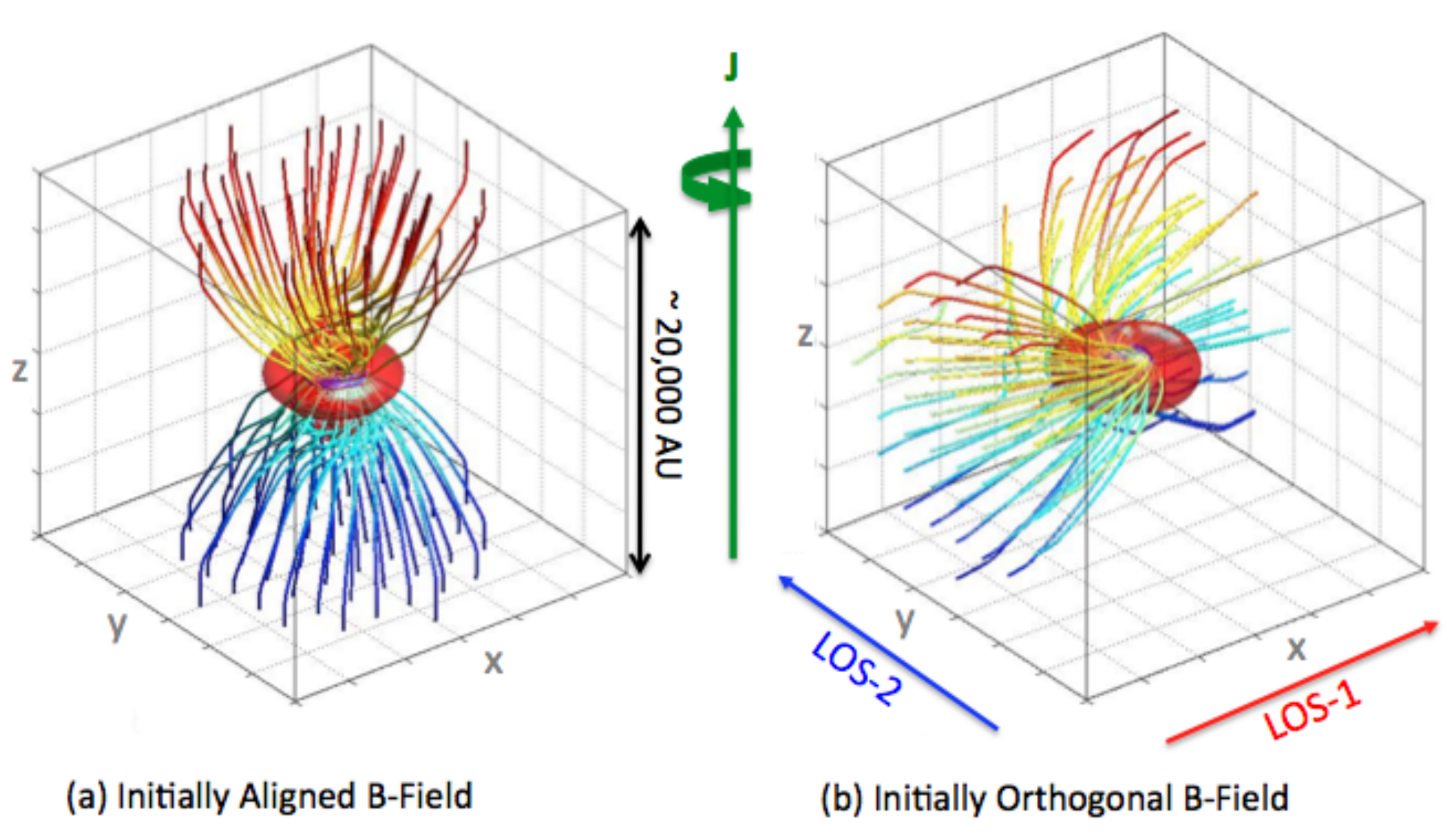}
\caption{\small{
3-D views of the magnetic field lines in the early stages of a global cloud core collapse with rotation ($\bold{J}$), before the field lines become very twisted. (a) shows a collapse where the B-field was initially aligned with the rotation of the cloud (e.g., the SA and WA models), while (b) shows a collapse where the field lines were initially orthogonal to the rotation of the cloud (e.g. the SO and WO models). The two lines of sight, LOS-1 and LOS-2, defined here are very different for the orthogonal SO and WO models, and these are explicitly examined throughout this paper. For L1527, the rotation axis lies in the plane of the sky in the east-west direction.}}  
\end{center}
\end{figure}

\begin{figure}[p]
\begin{center}
\includegraphics[width=4in]{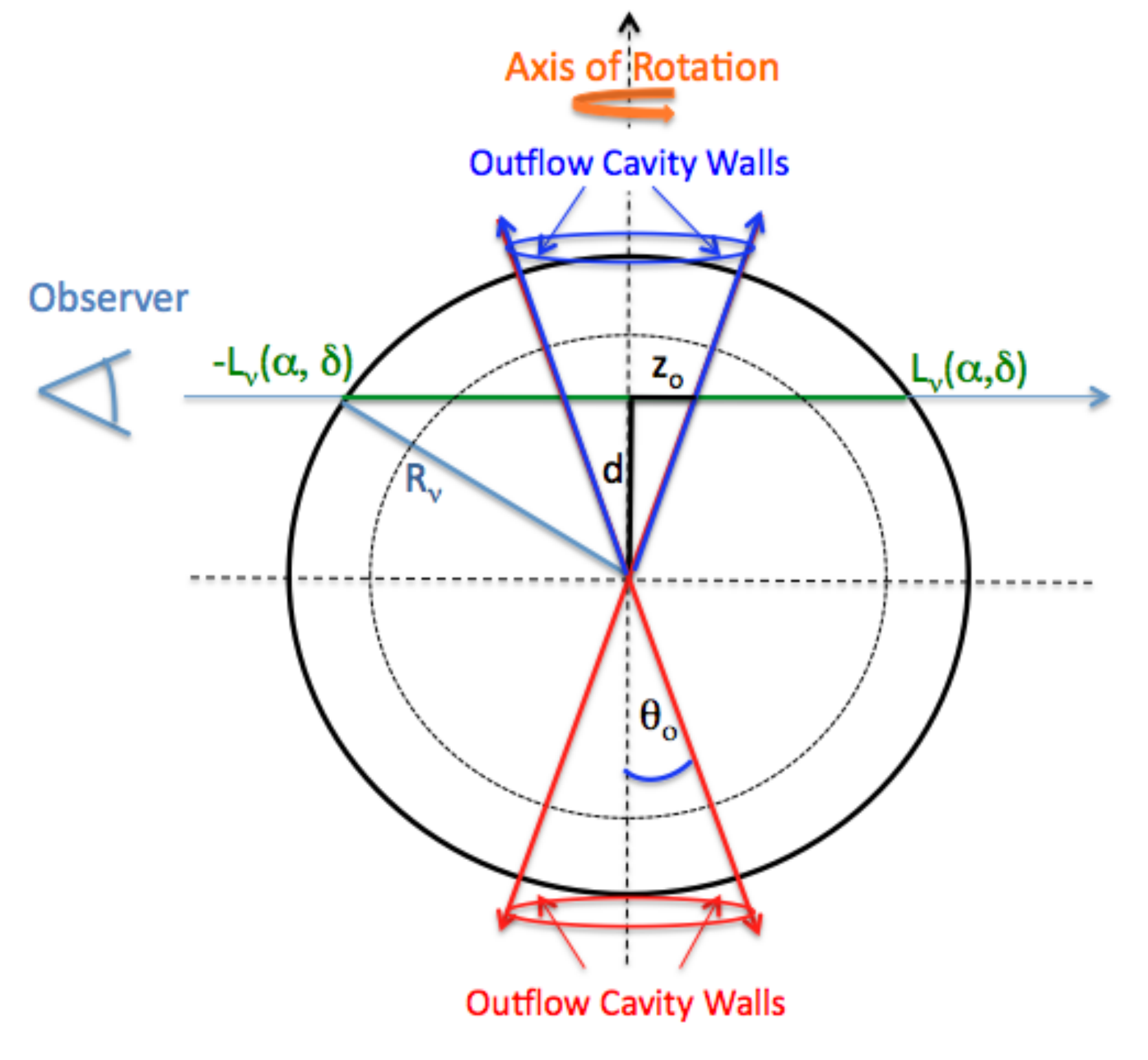}
\caption{\small{
Illustrates the geometry used to define the line of sight limits for each observation set. This includes a simple model of the outflow cavities as two back-to-back cones with half-opening angles of $\theta_o$.  Note: we have assumed for this model that the axis of the outflow is aligned with the axis of rotation, and that both lie in the plane of the sky aligned with the E-W axis (i.e., the $\alpha$ axis). Quantities L$_{\nu}$, R$_{\nu}$, d, and z$_o$ are defined in Section 2. }}  
\end{center}
\end{figure}

\begin{sidewaysfigure}[p]
\begin{center}
\includegraphics[height=5.5in]{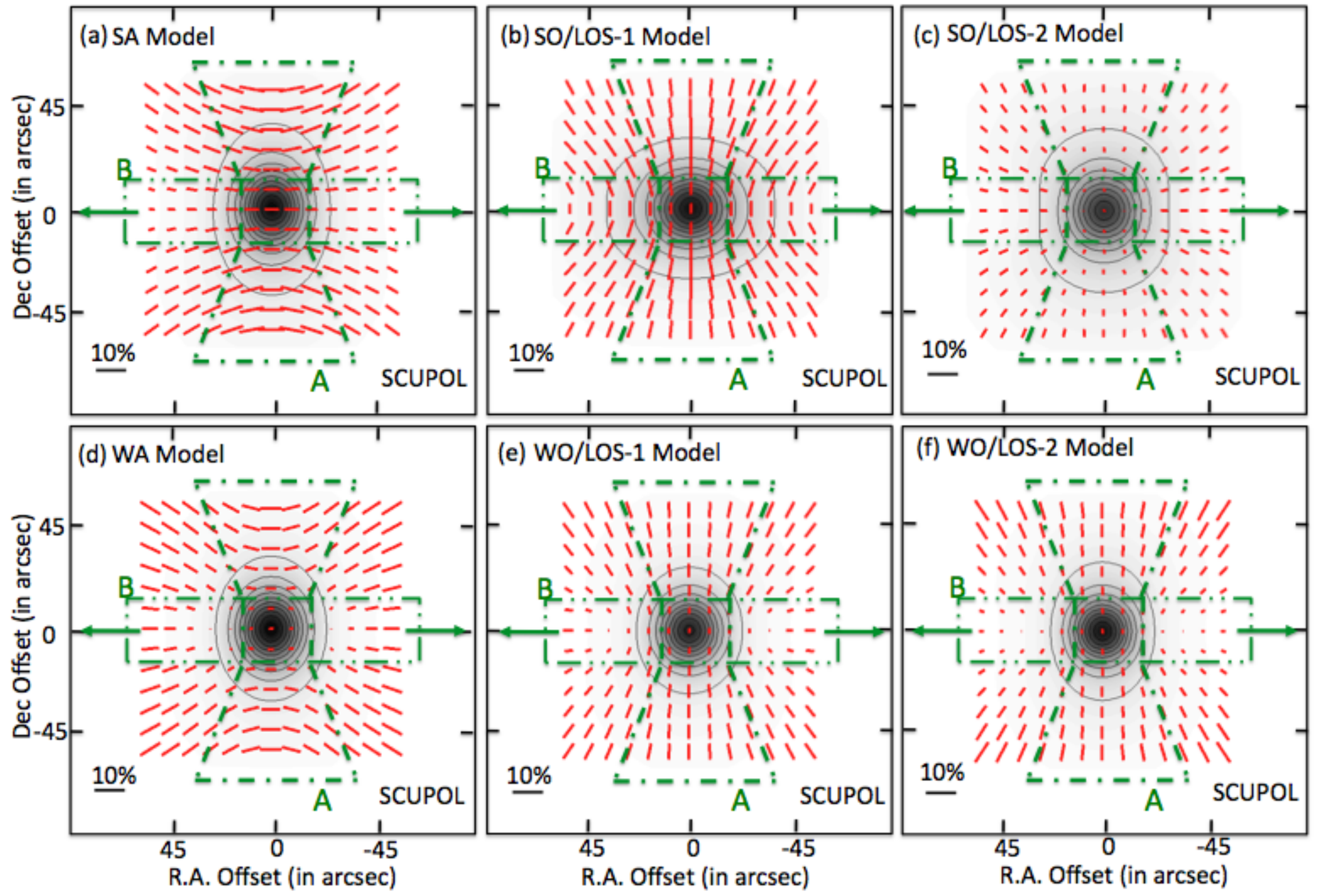}
\caption{ \footnotesize{
The \underline{model} B-vectors (in red) for the L1527 SCUPOL observation set. The model used is labelled on each B-vector map.  The rotation and outflow axes are oriented E-W (green arrows).  The vectors are superimposed on the contour/greyscale intensity maps of the corresponding model, convolved with a SCUPOL beam. See Section 3 of the text for a full explanation of this figure.
}}  
\end{center}
\end{sidewaysfigure}

\begin{sidewaysfigure}[p]
\begin{center}
\includegraphics[height=5.5in]{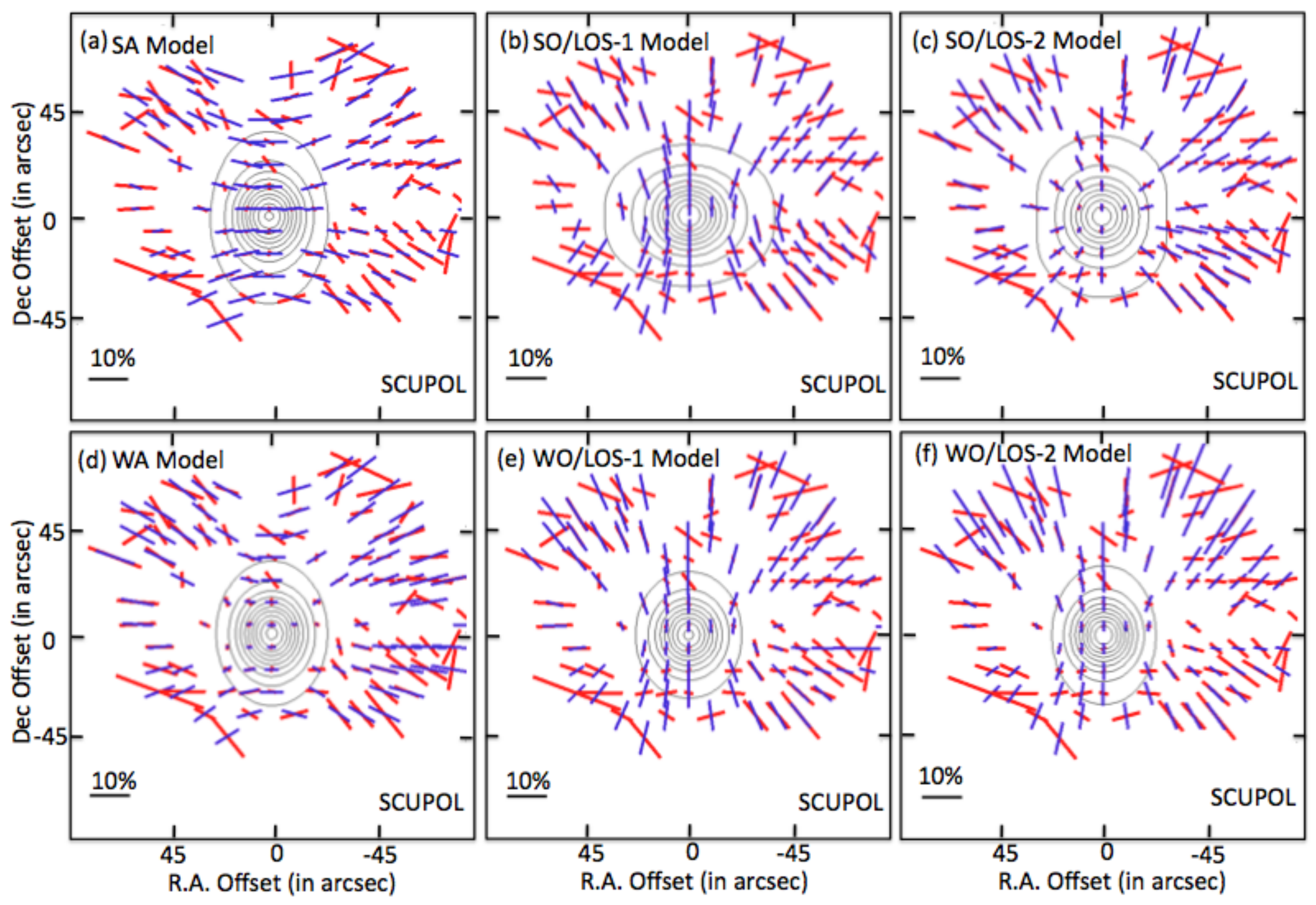}
\caption{\footnotesize{
Comparison between model SCUPOL B-vectors (blue) and the observed SCUPOL B-vectors (red) in L1527. The model used is labelled on each B-vector map. The rotation and outflow axes are oriented E-W. The vectors are superimposed on the contour maps of the model intensity, convolved with a SCUPOL beam. These contours are in increments of one-tenth of the peak.}}   
\end{center}
\end{sidewaysfigure}

\begin{sidewaysfigure}[p]
\begin{center}
\includegraphics[height=5.5in]{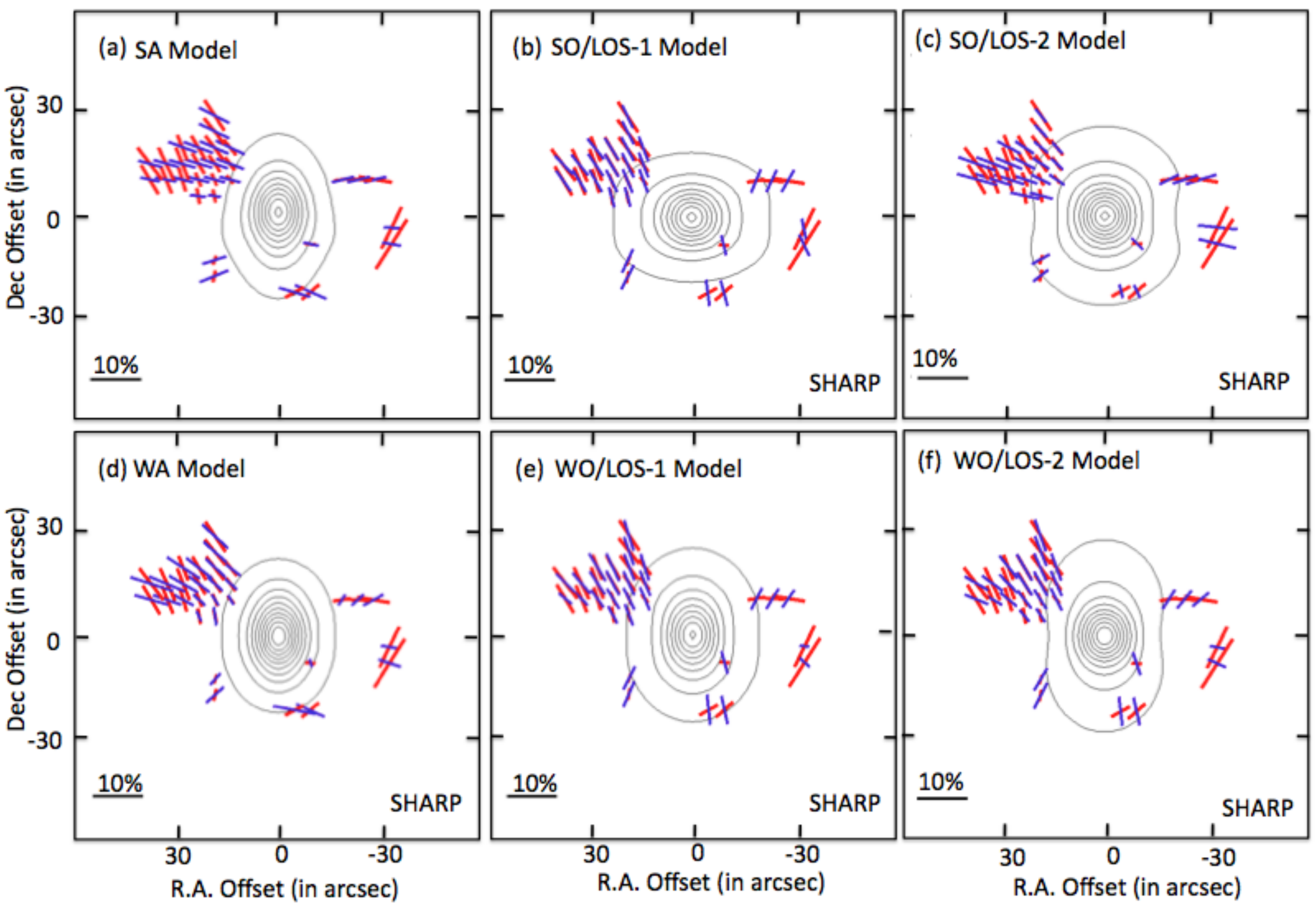}
\caption{\footnotesize{
Comparison between model SHARP B-vectors (blue) and the observed SHARP B-vectors (red) in L1527. The model used is labelled on each B-vector map. The rotation and outflow axes are oriented E-W. The vectors are superimposed on the contour maps of the model intensity, convolved with a SHARP beam. These contours are in increments of one-tenth of the peak.}}
\end{center}
\end{sidewaysfigure}

\begin{sidewaysfigure}[p]
\begin{center}
\includegraphics[height=5.5in]{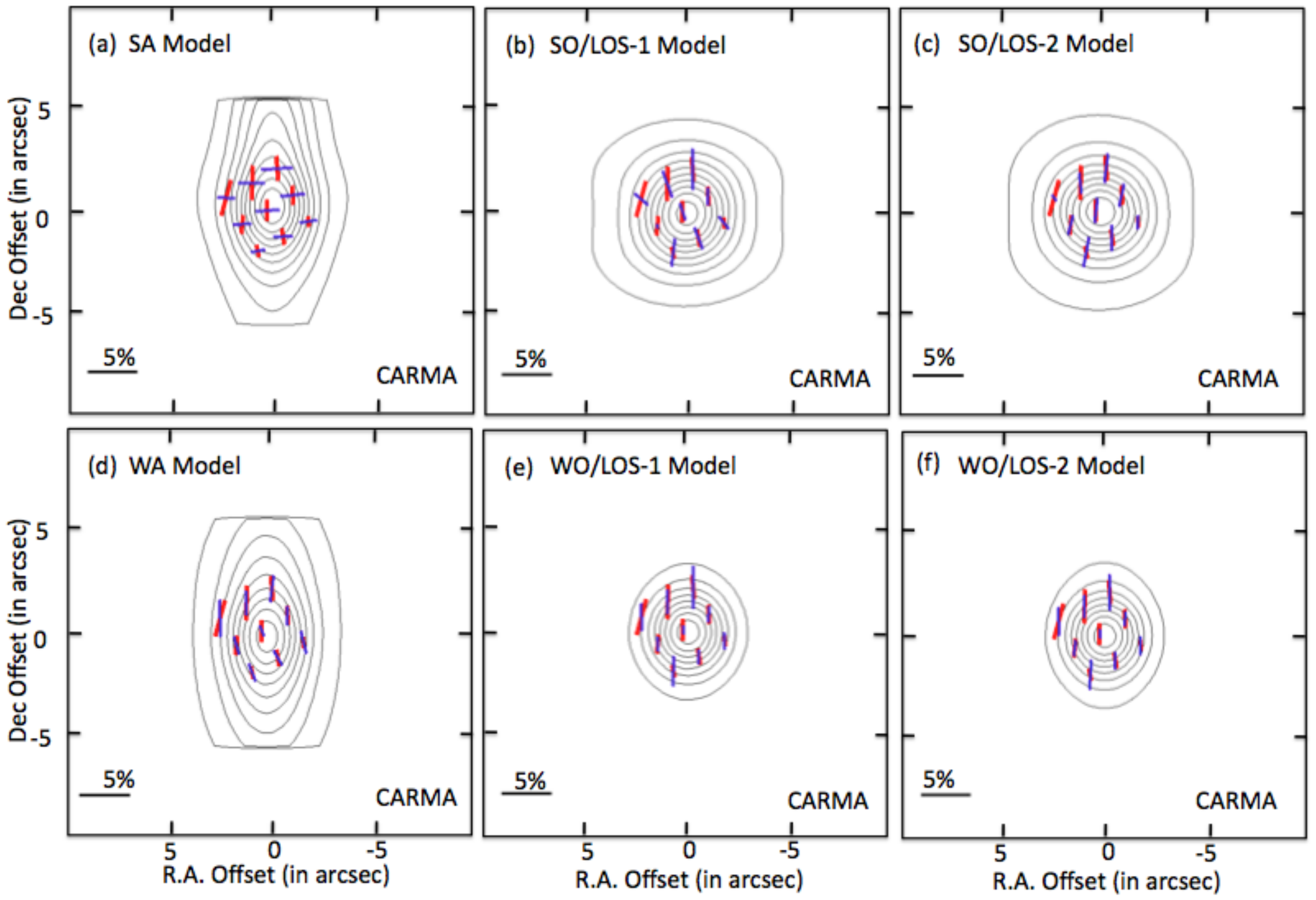}
\caption{\footnotesize{
Comparison between model CARMA B-vectors (blue) and the observed CARMA B-vectors (red) in L1527. The model used is labelled on each B-vector map. The rotation and outflow axes are oriented E-W. The vectors are superimposed on the contour maps of the model intensity, convolved with a SHARP beam. These contours are in increments of one-tenth of the peak.}}
\end{center}
\end{sidewaysfigure}

\bibliography{sample}

\begin{thebibliography}{64}
\providecommand{\natexlab}[1]{#1}

\bibitem[]{} Allen, A., Shu, F.H., \& Li, Z.-Y. 2003, \apj, 599, 351 
\bibitem[]{} Allen, A., Li, Z.-Y., \& Shu, F.H. 2003, \apj, 599, 363
\bibitem[]{} Basu, S. \& Mouschovias, T.C. 1994 \apj, 432, 720
\bibitem[]{} Benson, P.J. \& Myers, P.C. 1989, \apjs, 71, 89
\bibitem[]{} Bouvier, J., Cabrit, C., Ferna«ndez, M., Martõ«n, E. L., \& Matthews, J. 1993, \aap, 272, 176
\bibitem[]{} Chandler, C.J., \& Richer, J.S. 2000, \apj, 530, 851 
\bibitem[]{} Chapman, N. L., Davidson, J. A., Goldsmith, P. F., et al. 2013, \apj, 770, 151
\bibitem[]{} Chapman, N.L., Goldsmith, P.F., Pineda, J.L., Clemens, D.P., Li, D., \& Krco M. 2011, \apj, 741, 21
\bibitem[]{} Crutcher, R.M. 1999, \apj, 520, 706
\bibitem[]{} Dapp, W.B., \& Basu, S. 2010, \aap, 521, L56
\bibitem[]{} Davidson, J. A., Novak, G., Matthews, T. G., et al. 2011, \apj, 732, 97
\bibitem[]{} Dennison, B. 1977, \apj, 215, 529
\bibitem[]{} Dotson, J. L., Vaillancourt, J. E., Kirby, L., et al. 2010, \apjs, 186, 406
\bibitem[]{} Galli, D., Lizano, S., Shu, F. H., \& Allen, A. 2006, \apj, 647, 374
\bibitem[]{} Goodman, A. A., Jones, T. J., Lada, E. A., \& Myers, P. C. 1995, \apj, 448, 748
\bibitem[]{} Goodman, A.A., Benson, P.J., Fuller, G.A., \& Myers, P.C. 1993, \apj, 406, 528
\bibitem[]{} Goodman, A. A., Jones, T. J., Lada, E. A., \& Myers, P. C. 1992, \apj, 399, 108
\bibitem[]{} Greaves, J. S., et al. 2003, \mnras, 340, 353
\bibitem[]{} Hennebelle, P.  \& Ciardi, A. 2009. \aap, 506, L29
\bibitem[]{} Hildebrand, R.H., Davidson, J.A., Dotson, J.L., Dowell, C.D., Novak, G., \& Vaillancourt, J.E. 2000, \pasp, 112, 1215
\bibitem[]{} Hildebrand, R.H., \& Dragovan, M. 1995, \apj, 450, 663
\bibitem[]{} Hildebrand, R.H. 1988, \qjras, 29, 327
\bibitem[]{} Hogerheijde, M.R, van Dishoeck, E.F, Blake, G.A., \& van Langevelde, H.J. 1998, \apj, 502, 315
\bibitem[]{} Hull, C. L. H., Plambeck, R. L., Kwon, W., et al. 2014, \apjs, 213, 13
\bibitem[]{} Hull, C. L. H., Plambeck, R. L., Bolatto, A. D., et al. 2013, \apj, 768, 159
\bibitem[]{} Jones, T.J., Bagley, M., Krejny, M., Andersson, B-G, \& Bastien, P. 2014 (submitted to \aj)
\bibitem[]{} Jones, T.J. 1989, \apj, 346, 728
\bibitem[]{} Jones, T.W., \& O'Dell, S.L. 1977, \apj, 214, 522
\bibitem[]{} Joos, M., Hennebelle, P., \& Ciardi, A. 2012, \aap, 543, 128
\bibitem[]{} J\o rgensen, J. K., van Dishoeck, E. F., Visser, R., et al. 2009, \aap, 507, 861
\bibitem[]{} Kenyon, S. J., Dobrzycka, D., \& Hartmann, L. 1994, \aj, 108, 1872
\bibitem[]{} Krumholz, M.R., Crutcher, R.M., \& Hull, C.L.H. 2013, \apj, 767, L11
\bibitem[]{} Kwon, W., Looney, L.W., Mundy, L.G., Chiang, H-F., \& Kemball, A.J. 2009, \apj, 696, 841
\bibitem[]{} Ladd, E.F., Adams, F.C., Casey, S., Davidson, J.A., Fuller, G.A., Harper, D.A., Myers, P.C., \& Padman, R. 1991, \apj, 366, 203
\bibitem[]{} Lazrarian, A. 2007, \jqsrt, 106, 225
\bibitem[]{} Lee, H.M., \& Draine, B.T. 1985, \apj, 290, 211 
\bibitem[]{} Li, H., Dowell, C. D., Kirby, L., Novak, G., \& Vaillancourt, J. E. 2008, App. Opt., 47, 422
\bibitem[]{} Li, H., Griffin, G. S., Krejny, M., et al. 2006, \apj, 648, 340
\bibitem[]{} Li, Z-Y.,  Krasnopolsky, R., \& Shang, H. 2013. \apj, 774, 82
\bibitem[]{} Matthews, B.C., McPhee, C.A., Fissel, L.M., \& Curran, R.L. 2009, \apjs, 182, 143
\bibitem[]{} McKee, C.F. \& Ostriker, E.C. 2007, \araa, 45, 565
\bibitem[]{} Mouschovias, T.C. \& Paleologou, E. V. 1979 \apj, 230 204
\bibitem[]{} Mouschovias, T.C. \& Paleologou, E. V. 1980 \apj, 237, 877
\bibitem[]{} Myers, P.C., Bachiller, R., Caselli, P. , Fuller, G.A., Mardones, D., Tafalla, M.,  \& Wilner, D.J. 1995, \apj, 449, L65
\bibitem[]{} Nakano, T., \& Nakamura, T. 1978, \pasj,  30, 671
\bibitem[]{} Ohashi, N., Hayashi, M., Ho, P.T.P, \& Momose, M. 1997, \apj, 475, 211  
\bibitem[]{} Ossenkopf, V., \& Henning, T. 1994, \aap, 291, 943
\bibitem[]{} Santos-Lima, R., de Gouvela Dal Pino, E.M., \& Lazarian, A. 2012, \apj, 747, 21
\bibitem[]{} Shu, F.H. 1977, \apj, 214, 488
\bibitem[]{} Spitzer, L., Jr. 1978, Physical Processes in the Interstellar Medium (New York: Wiley)
\bibitem[]{} Stephens, I. W., Looney, L. W., Kwon, W., et al. 2013, \apj, 769, L15
\bibitem[]{} Tobin, J. J., Hartmann, L., Chiang, H.-F., et al. 2012, \nat, 492, 83
\bibitem[]{} Tobin, J. J., Hartmann, L., Chiang, H-F., et al.  2011 \apj 740, 45 
\bibitem[]{} Tobin, J. J., Hartmann, L. \& Loinard, L.  2010 \apj L. 722, L12
\bibitem[]{} Tobin, J. J., Hartmann, L., Calvet, N. \& D'Alessio, P. 2008,  \apj, 679, 1364
\bibitem[]{} Tomisaka, K. 1998, \apj, 502, L163
\bibitem[]{} Tomisaka, K.  2000, \apj, 528, L41
\bibitem[]{} Torres, R.M., Loinard, L., Mioduszewski, A.J., \& Rodriquez, L.F. 2009, \apj, 698, 242
\bibitem[]{} Troland, T. H., \& Crutcher, R. M. 2008, \apj, 680, 457
\bibitem[]{} Vaillancourt, J.E., \& Matthews, B.C. 2012, \apjs, 201, 13
\bibitem[]{} Whittet, D. C. B., Hough, J. H., Lazarian, A., \& Hoang, T. 2008, \apj, 674, 304
\bibitem[]{} Williams, J. P., \& Cieza, L. A. 2011, \araa, 49, 67
\bibitem[]{} Zeng, L., Bennett, C.L., Chapman, N.L., Chuss, D.T., Jimenez-Serra, I., Novak, G., \& Vaillancourt, J.E. 2013, \apj, 773, 29
\bibitem[]{} Zhou, S., Evans, N.J., II, \& Wang, Y. 1996, \apj, 466, 296 

\end{thebibliography}

\end{document}